\documentclass[twocolumn]{aastex62}

\usepackage[graphicx]{realboxes} 
\received{January 1, 2018}
\revised{January 7, 2018}
\accepted{\today}
\submitjournal{ApJ}

\shorttitle{Stellar abundances in NGC 1313}
\shortauthors{Hernandez et al.}

\begin{document}

\title{\uppercase{Chemical abundances of Young Massive Clusters in NGC 1313}\footnote{Based on observations made with ESO telescopes at the La Silla Paranal Observatory under program ID 084.B-0468(A).}}

\correspondingauthor{Svea Hernandez}
\email{sveash@stsci.edu}

\author[0000-0003-4857-8699]{Svea Hernandez}
\affiliation{AURA for ESA, Space Telescope Science Institute, 3700 San Martin Drive, Baltimore, MD 21218, USA}

\author{Autumn Winch}
\affiliation{Bryn Mawr College,
101 N Merion Ave, 
Bryn Mawr, PA 19010, USA}

\author{S{\o}ren Larsen}
\affiliation{Department of Astrophysics / IMAPP,
Radboud University, 
PO Box 9010, 
6500 GL Nijmegen, The Netherlands}

\author{Bethan L. James}
\affiliation{AURA for ESA, Space Telescope Science Institute, 3700 San Martin Drive, Baltimore, MD 21218, USA}

\author{Logan Jones}
\affiliation{Space Telescope Science Institute,
3700 San Martin Drive, 
Baltimore, MD 21218, USA}

\begin{abstract}
We analyze spectroscopic observations of five young massive clusters (YMCs) in the barred spiral galaxy NGC 1313 to obtain detailed abundances from their integrated light. Our sample of YMCs was observed with the X-Shooter spectrograph on the Very Large Telescope (VLT). We make use of theoretical isochrones to generate synthetic integrated-light spectra, iterating on the individual elemental abundances until converging on the best fit to the observations. We measure abundance ratios for [Ca/Fe], [Ti/Fe], [Mg/Fe], [Cr/Fe], and [Ni/Fe]. We estimate an Fe abundance gradient of $-$0.124 $\pm$ 0.034 dex kpc$^{-1}$, and a slightly shallower $\alpha$ gradient of $-$0.093 $\pm$ 0.009 dex kpc$^{-1}$. This is in contrast to previous metallicity studies that focused on the gas-phase abundances, which have found NGC 1313 to be the highest-mass barred galaxy known \textit{not} to have a radial abundance gradient. We propose that the gradient discrepancy between the different studies originates from the metallicity calibrations used to study the gas-phase abundances. We also observe an age-metallicity trend which supports a scenario of constant star formation throughout the galaxy, with a possible burst in star formation in the south-west region where YMC NGC 1313-379 is located. 
 \end{abstract}


\section{Introduction}\label{sec:intro}
Understanding how galaxies evolve chemically continues to be a fundamentally important topic in modern astrophysics. Stellar abundances are emerging as powerful tools for tracing the chemical composition of Galactic and extragalactic stellar populations, as the signatures of the gas reservoirs that formed such populations are preserved in their  atmospheres. Through the analysis of abundance patterns of individual elements, particularly those which are sensitive to the different timescales of galactic evolution, one can reliably trace the enrichment history of individual galaxies. For example, the ratio of $\alpha$ elements (i.e. Ca, Ti, Mg, Si) to Fe-peak elements (i.e. Fe, Mn, Cr, Sc) can be used as indicator of the relative contribution from core-collapsed supernovae (SNe) and type Ia SNe \citep{tin79,mcw97}. \par
In the last decades, chemical abundance studies of individual stars in the Milky Way (MW) have provided us with an extremely detailed picture of various Galactic components. The analysis of high-dispersion spectroscopy of individual stars in our Galaxy have shown that halo and bulge stars are primarily old with enhanced abundance ratios of $\alpha$ over Fe \citep{sne79,ful00}. Stars in the disk, on the other hand, are observed to be young with abundance trends similar to those observed in the sun \citep{edv93,ben03,red03,ben14}. A natural next step is to perform similarly detailed studies on individual stars outside of the MW, and beyond the Local Group.\par
In star-forming galaxies, most of the abundance work is done through the analysis of emission lines from \ion{H}{2} regions \citep{sea71}. The gas-phase  abundances obtained from these objects, however, do not usually provide constraints on ratios such as [$\alpha$/Fe]. An alternative approach to characterize the chemical patterns in extragalactic environments is the spectroscopic technique developed in the last couple decades, relying on both red \citep[RSG,][]{dav10} and blue supergiants \citep[BSG,][]{bre06,bre16,eva07}. Such studies have shown that accurate metallicities can be obtained through the analysis of spectroscopic observations from RSGs/BSGs in agreement ($\sim$0.1-0.3 dex) with those inferred from \ion{H}{2} regions \citep{bre09, dav17}. \par
Another promising probe of detailed abundances in extragalactic environments are star clusters. Since the spectra of the integrated light (IL) of stellar clusters are broadened by a few km s$^{-1}$, the individual stellar absorption lines are easily resolved. High-resolution (R $\sim$ 30,000) spectra of globular clusters (GCs) have been reliably used to study the chemical histories of the MW and nearby galaxies \citep{ber02, mcw08, col09,col11,col12,col17, lar12,lar17, lar18a, lar18b}. More recently the analysis of the IL of star clusters was expanded not only to younger stellar populations than GCs \citep{lar06,lar08} but also to lower resolution observations \citep[R $\lesssim$ 8,000,][]{gaz14,lard15,her17,her18}. The work presented here aims at exploiting these recently developed IL analysis tools to dissect the chemical history of the nearby barred spiral galaxy, NGC 1313.\par
\begin{figure*}
   	  \centerline{\includegraphics[scale=0.43]{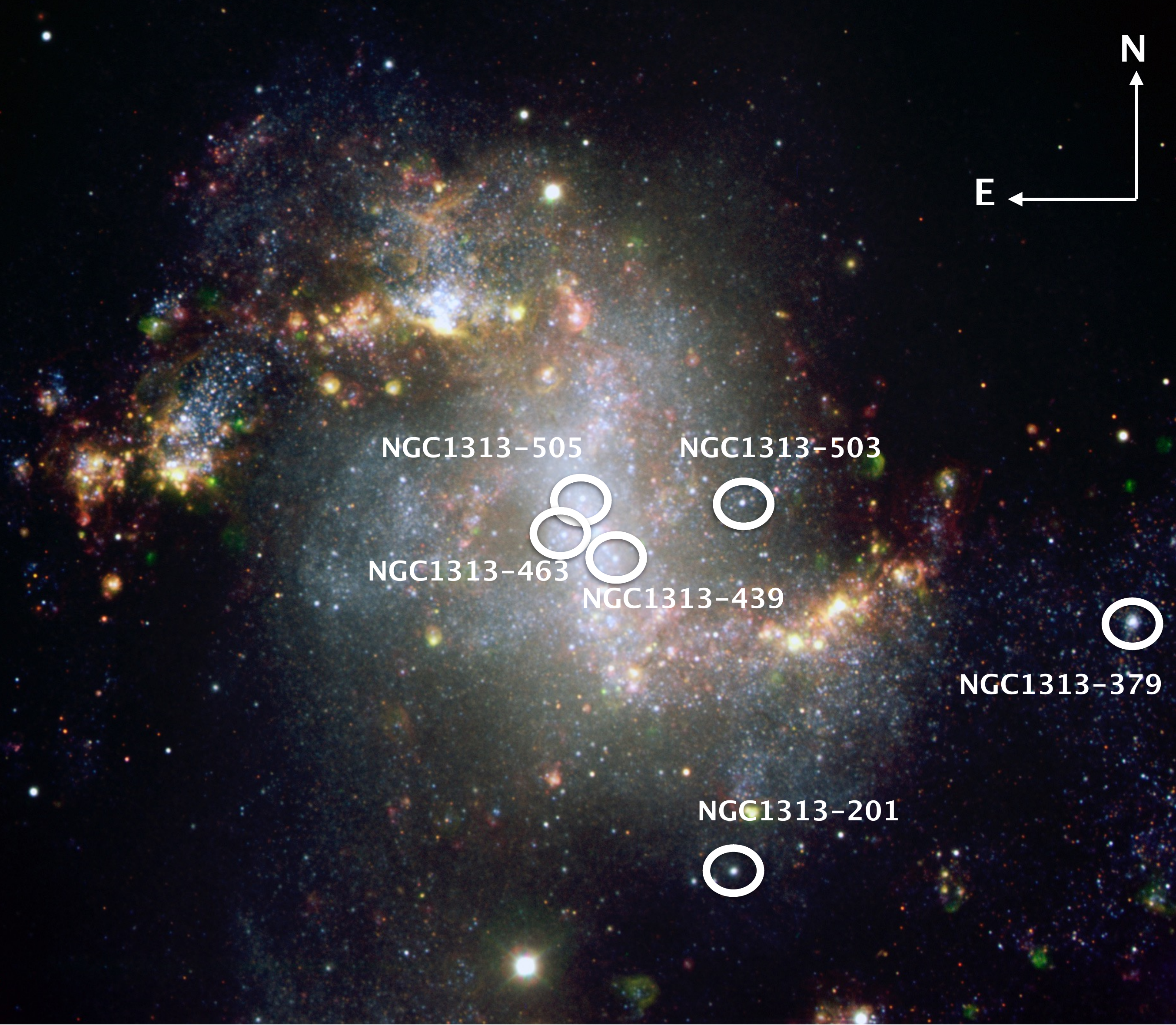}}
      \caption{Color-composite image observed with the FORS1 instrument on the ESO's Very Large Telescope (VLT). The observations were downloaded from the ESO Science Archive and fully processed by Henri Boffin (ESO). Credit: ESO. The white apertures show the locations of the YMCs studied here, along with NGC1313-379 from \citet{her17}.}
         \label{fig:ngc1313}
   \end{figure*}

At a distance of 4.4 Mpc \citep{Jac09}, NGC 1313 has caught the attention of many studies due to its peculiar morphology (Figure \ref{fig:ngc1313}), possibly indicating an interaction with a satellite companion \citep{san79,bla81}. \citet{lar07} and \citet{sil12} have concluded that there is a particular increase in star formation activity in the south-west region of this galaxy supporting tidal interaction scenarios previously proposed by \citet{pet94} and \citet{bla81}. Interestingly, one of the most massive star clusters, NGC1313-379, is located in this precise region and was previously studied by \citet{her17}. The detailed abundances of this YMC showed close-to-solar [$\alpha$/Fe] = $+$0.06 $\pm$ 0.11 dex, also similar to the nebular metallicities measured for two nearby \ion{H}{2} regions \citep{wal97}. We list in Table \ref{table:gal} some of the properties of this nearby galaxy. \par 
Our analysis aims at expanding the exploratory  work of \citet{her17} by adding five more YMCs distributed throughout NGC 1313. Using the IL analysis tools first developed by \citet{lar12} along with observations taken with the X-Shooter spectrograph on the Very Large Telescope (VLT), we perform a detailed abundance study of the young stellar populations in this nearby barred spiral galaxy. The contents of this paper are structured as follows: Section \ref{sec:obs} describes the spectroscopic observations and the data reduction, Section \ref{sec:ana} details the analysis done, in Section \ref{sec:results} we present our results, and Sections \ref{sec:discussion} and \ref{sec:con} present our findings and our conclusions. \par

\begin{table}
\caption{Properties of NGC 1313}
\label{table:gal}
\centering 
\begin{tabular}{lc}
\hline \hline
Parameter & Value \\
\hline\\
R. A. (J2000.0) & 49.566875\\
Dec (J2000.0) & -66.498250\\
Distance$^{a}$ & 4.25 Mpc\\
Morphological type & SB(s)d \\
$R_{25}\:^{b}$ & 4.\arcmin56 (6.02 kpc)\\
Inclination$^{c}$ & 48$^{\circ}$ \\
Position Angle$^{c}$ & 0$^{\circ}$\\
Heliocentric radial velocity & 470 km s$^{-1}$\\
\hline
\end{tabular}
\begin{minipage}{15cm}~\\
\textbf{Notes.} All parameters from the NASA Extragalactic \\Database (NED), except where noted.\\
 \textsuperscript{$a$}{\citep{tul16}} \\
 \textsuperscript{$b$}{\citep{vau91}} \\
 \textsuperscript{$c$}{\citep{ryd95}}\\
 \end{minipage}
\end{table}


\section{Observations and Data Reduction}\label{sec:obs}
The analysis presented here relies on X-Shooter  spectroscopic observations taken as part of the VLT guaranteed time observation (GTO) program ID: 084.B-0468A. The data were collected on November 2009, executing in standard nodding mode with an ABBA sequence. The X-Shooter spectrograph provides observations spanning a wavelength coverage ranging between 3000 and 24800 \r{A} at resolving power of $R=$3000--17000, depending on the configuration used. This instrument makes use of a three-arm system to simultaneously observe in three bands, UVB (3000--5600 \r{A}), VIS (5500--10200 \r{A}), and NIR (10200--24800 \r{A}). Our observations use slit widths of 1.0\arcsec, 0.9\arcsec, and 0.9\arcsec obtaining resolutions of $R\sim$ 5100, 8800, and 5100 for the UVB, VIS, and NIR arms, respectively. \par
The YMCs in our sample were primarily chosen from a cluster compilation by \citet{lar04}. The selection sample was focused on isolated targets. We highlight the location of the YMCs studied here in Figure ~\ref{fig:ngc1313}, also including the cluster studied in \citet[][NGC1313-379]{her17}. We list in Table \ref{table:ymc_prop} the physical properties of the clusters in our sample. In Table \ref{table:obs} we present information on the X-Shooter observations including exposure times used for the different arms, exposure dates, and signal-to-noise (S/N) ratios for the individual arms. We point out that given the low S/N ratios observed in the NIR data we based the analysis presented here solely on the UVB and VIS observations.\par
The observations were calibrated using the X-Shooter pipeline (v.2.5.2) with the European Southern Observatory (ESO) Recipe Execution Tool (ESOREX) v.2.11.1. The software allowed us to perform the basic data reduction steps including dark and bias corrections, flat-fielding, wavelength calibration and sky subtraction. Similar to the approach in \citet{her17}, we make use of the IDL routines by \citet{che14} to extract the science spectrum from the calibrated 2D images and combine the different orders using a variance-weighted average in the overlapping regions. We flux calibrate the extracted 1D spectra using the spectrophotometric standard Feige 110. Furthermore, we create response curves for each of the Feige 110 exposures using the same bias and master flats as those used in the reduction of the science observations, correcting for exposure time and atmospheric extinction. As highlighted in \citet{her18}, this last step of applying the same master flats in both the science and the response curves is critical to remove any contemporaneous flat-field features. \par
Telluric contamination from the Earth's atmosphere strongly affect the X-Shooter observations particularly the data from the VIS and NIR arm. We used the routines created by \citet{che14} to remove any contribution from the telluric absorption features. The routines by \citet{che14} are based on Principal Component Analysis (PCA) which rely on a telluric library of 152 spectra to remove and reconstruct the strongest telluric absorption. We highlight that although we apply the PCA technique by \citet{che14} to correct for this contaminating features, throughout our analysis we aim at avoiding these strongly affected regions: 5800--6100 \r{A}, 6800--7400 \r{A}, 7550--7750\r{A}, 7800--8500 \r{A}, 8850--10000 \r{A}.\par
\begin{figure*}
   
   	  \centerline{\includegraphics[scale=0.7]{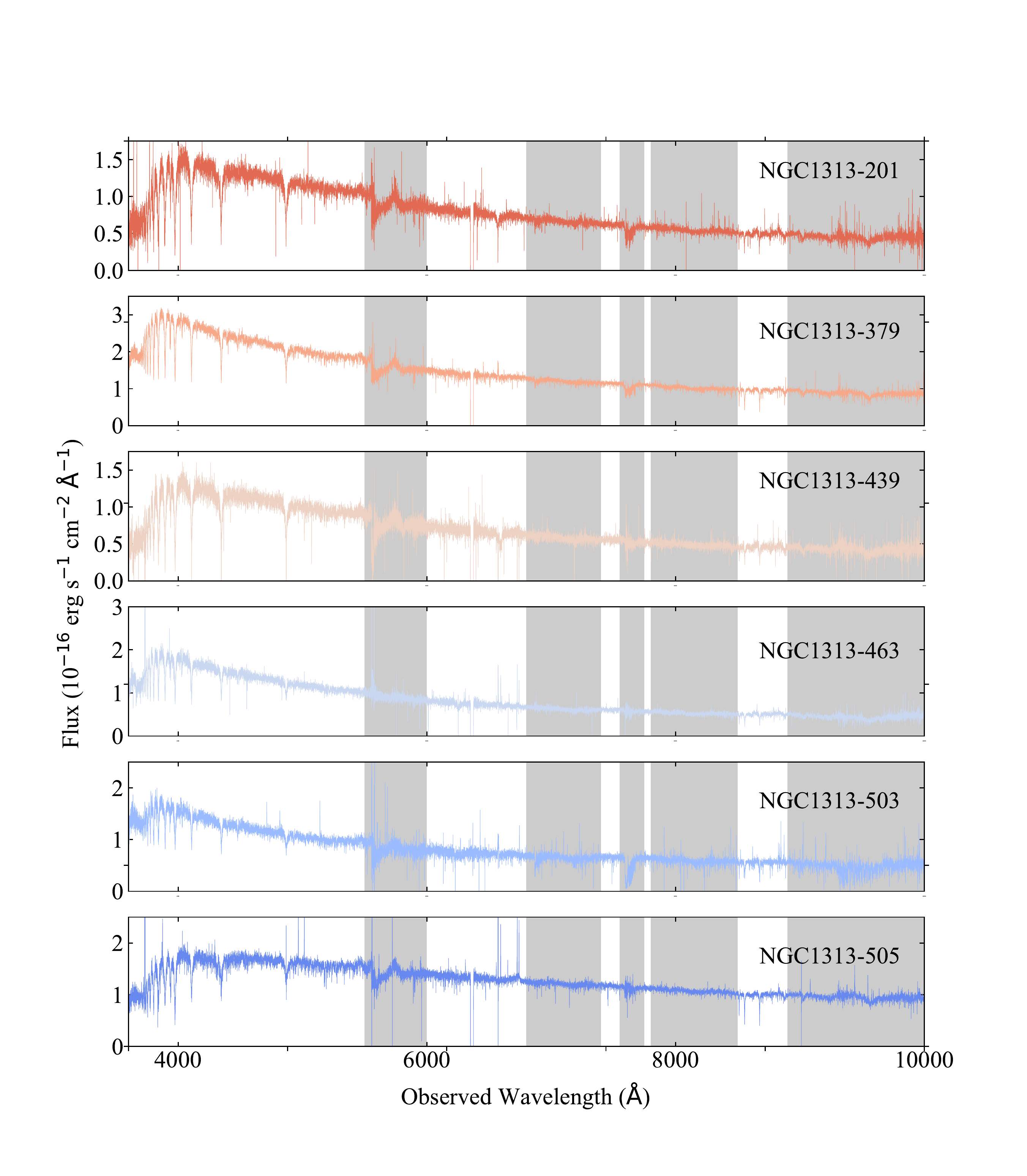}}
      \caption{Fully calibrated X-Shooter spectra for a sample of YMCs in NGC 1313. Our analysis focuses on the UVB and VIS X-Shooter arms, spanning a wavelength coverage of 3000 to 10,000 $\rm \AA$. Note the strong dichroic feature around 5700 $\rm \AA$ observed in the VIS arm. We show in grey the regions contaminated by telluric absorption and dichroic features. These regions have been excluded from the fitting procedure.}
         \label{fig:ngc1313_spectra}
   \end{figure*}

We show in Figure \ref{fig:ngc1313_spectra} the individual spectra for each of our targets. We note that in all of our clusters we observe strong dichroic features around 5700 $\rm \AA$. According to \citet{che14} these features typically appear in the 1D spectra in slightly different locations, which makes removing them a complicated task. Similar to those wavelength regions affected by telluric contamination, we also exclude from our abundance analysis those wavelength ranges affected by this instrumental feature. 

\begin{table}
\caption{Physical properties of the YMC sample studied here.}
\label{table:ymc_prop}
\centering 
\begin{tabular}{lcccc}
\hline \hline
Cluster & RA & Dec &Age$^b$ &R/R$_{\rm 25}$ \\
 & (J2000) & (J2000) & (Myr)& \\
\hline\\
NGC1313-201 &	 49.536535&	 $-$66.52634&	 300& 0.44\\
NGC1313-379$^{a}$&	 49.449038&	 $-$66.50474& 55& 0.92\\
NGC1313-439 & 49.561676&	 $-$66.49921& 200& 0.04\\
NGC1313-463 & 49.572854&	 $-$66.49715& 40& 0.05\\
NGC1313-503 & 49.536164&	 $-$66.49457& 20& 0.24\\
NGC1313-505 & 49.571406&	 $-$66.49454& 100& 0.06\\
\hline
\end{tabular}
\begin{minipage}{15cm}~\\
 \textsuperscript{$a$}{Analysis done in \citet{her17}}\\
  \textsuperscript{$b$}{Taken from the LEGUS cluster catalog \citep{ada17}}\\
 \end{minipage}
\end{table}

\begin{table*}
\caption{Observed YMCs and their X-Shooter observations}
\label{table:obs}
\centering 
\begin{tabular}{cccccccc}
\hline \hline
Cluster & \multicolumn{3}{c}{$t_{\rm exp}$ (s)} & Date &  \multicolumn{3}{c}{S/N (pix$^{-1}$)} \\
 & UVB & VIS & NIR &  &  UVB & VIS & NIR\\
\hline\\	
NGC1313-201&	 1620.0&	 1600.0&	 1680.0&	 2009-11-25T07:13:21.8922&	  23.6&	 18.1	& 4.0\\	
NGC1313-379$^{a}$&	 3120.0&	 3080.0&	 3240.0&	 2009-11-25T04:00:15.9896&	  46.3&	 39.4	& 13.4\\
NGC1313-439& 1620.0&	 1600.0&	 1680.0&	 2009-11-25T06:35:17.2808&	  16.1&	 12.2	& 4.2\\	
NGC1313-463&2260.0&	 2400.0&	 860.0&	 2009-11-22T06:23:33.3859&	  20.4&	 15.2	& 5.7\\	
NGC1313-503& 1620.0&	 1600.0&	 1680.0&	 2009-11-25T07:47:40.8959&	  20.4&	 16.7	& 7.1\\	
NGC1313-505& 3120.0&	 3080.0&	 3240.0&	 2009-11-25T05:02:43.7440&	 30.2	& 31.5&	 12.3\\	
\hline
\end{tabular}
\begin{minipage}{15cm}~\\
 \textsuperscript{$a$}{Analysis done in \citet{her17}}\\
 \end{minipage}
\end{table*}

\section{Analysis}\label{sec:ana}

The work presented here makes use of the IL analysis tool ISPy3 \citep{lar20}. ISPy3 relies heavily on spectral synthesis and full spectral fitting. This technique was primarily developed to analyze high-resolution spectroscopic observations ($R\sim$ 40,000). However, it was recently shown that the same software can be used to obtain detailed abundances from observations at lower resolutions \citep[$R < 10,000$;][]{her17,her18, her18b} using predefined wavelength windows which minimize line blending.\par
Briefly described, our analysis involve an iterative process where the abundances are modified each iteration until the best synthetic model is obtained for a given spectroscopic observation. This is done executing four main steps: 1) Derivation of stellar parameters for each of the stars in the clusters,  2) Creation of stellar atmospheric models, 3) Creation of IL synthetic spectrum, 4) Comparison between the modeled spectrum and the spectroscopic observations. In the following sections we expand on each of these steps. \par

\subsection{Stellar Parameters}\label{sec:stel_par}
The ISPy3 suite of Python routines rely on the information provided by the Hertzsprung–Russell diagram (HRD) of the star clusters under analysis. Three main approaches have been adopted for deriving the physical parameters from the HRD: [a] Color Magnitude Diagrams \citep[CMD;][]{lar17}, [b] CMD+Isochrones \citep{lar12,lar18,her17}, [c] Isochrones \citep{her18,her18b, her19}. Method [a] is optimally applied for nearby or local stellar clusters where resolved CMDs are available covering all evolutionary stages present. Method [b] is commonly used for extragalactic stellar clusters where the resolved CMDs typically include only the brightest stars in the cluster. The HRD is then complemented with theoretical isochrones covering fainter magnitudes. In the analysis presented here, we generate HRDs using solar-scaled stellar isochrones from PADOVA v.1.2.S \citep{bre12}, adopting the Isochrone only approach, [c]. The main driver for adopting this approach is the absence of CMDs for the YMCs in our sample. We note, however, that \citet{her17}  found that the differences between methods [b] and [c] have only a minor effect on the inferred abundances of YMC NGC 1313-379, of the order of $\lesssim$ 0.1 dex. \par
For each of the YMCs, we select the initial isochrones assuming a Large Magellanic Cloud (LMC)-like metallicity \citep{wal97}, [Z] = $-0.3$ dex, and the ages listed in Table \ref{table:ymc_prop} published in the Legacy Extragalactic UV Survey (LEGUS) cluster catalog \citep{ada17}. From the corresponding isochrones we extract the stellar parameters, e.g. effective temperatures (T$_{\rm eff}$), stellar masses (M), and surface gravities (log $g$), following a \citet{sal55} initial mass function (IMF) and a lower mass limit of 0.4 M$_{\odot}$. We note that \citet{her19} found that changing the choice of IMF (e.g. from Salpeter to Kroupa IMF) modifies the inferred metallicities on average by $\ll$ 0.1 dex. Lastly, similar to previous IL studies of YMCs we adopt microturbulent velocity values ($v_t$) depending on the effective temperature of the stars: $v_t=$ 2 km s$^{-1}$ for stars with T$_{\rm eff} <$ 6000 K, $v_t=$ 4 km s$^{-1}$ for stars with 6000 $<$ T$_{\rm eff}$ $<$ 22,000 K, and $v_t=$ 8 km s$^{-1}$ for stars with T$_{\rm eff}$ $>$ 22,000 K.\par

\subsection{Stellar Atmospheres}\label{sec:stel_atm}
Using the parameters described above, we generate atmospheric models for each stellar type. Similar to the work by \citet{her18, her18b}, we make use of the local thermodynamic equilibrium (LTE) one-dimensional (1D) plane-parallel atmosphere modeling code ATLAS9 \citep{kur70} for any stars with T$_{\rm eff}>$ 5000 K. For stars with T$_{\rm eff}<$ 5000 K we then use MARCS atmospheric models instead, as these have been primarily developed with a special focus on late-type stars \citep{gus08}. These MARCS models are LTE 1D plane-parallel or spherical. One important distinction between the two modeling codes is that we generate the ATLAS9 atmospheres on-the-fly, whereas the MARCS models have been precomputed and downloaded from their official website\footnote{\url{http://marcs.astro.uu.se}}. At this point in the analysis we set the initial abundances of the stellar population. We note that in the work presented here we adopt the Solar composition from \citet{gre98}.\par
We also highlight that the entirety of our analysis is based on LTE modelling. We currently do not apply any non-LTE (NLTE) corrections as these are dependent on the individual stellar types, and our work focuses on the integrated light of the star clusters. Previous studies have looked into the effects of NLTE treatment in the analysis of IL spectroscopic observations of GCs and found that the overall metallicities can change by $\sim$0.05 dex \citep{you19}. A different study by \citet{con17}, also focused on older stellar populations ($>$ 1 Gyr), concluded that the assumption of LTE is adequate when studying Na and Ca. Lastly, \citet{eit19} find that in many cases these NLTE corrections, specifically for Ba, Mg, and Mn, are significant ($>$ 0.1 dex) for metal-poor GCs. Most of these IL NLTE studies have so far focused on old populations of stars, where our focus here is on YMCs. 

\subsection{Synthetic Stellar Spectra}\label{sec:stel_spectra}
As it has been extensively described in our previous studies, our analysis requires the creation of individual synthetic spectra for each stellar type in the population under study. We continue to use the line lists by \citet{cas04}. For those atmospheres generated with ALTAS9 we use the SYNTHE software \citep{kur79,kur81}, and TURBOSPECTRUM \citep{ple12} for the MARCS atmospheric models. We then coadd the individual spectra to create a single IL model spectrum. \par
The IL model produced by the spectral synthesis codes is created at resolutions of $R\sim$ 500,000. To match the resolution of the X-Shooter data, $R\sim$ 5000--9000, we fit for the best Gaussian dispersion, $\sigma_{\rm sm}$, and convolve it with our IL synthetic spectrum. This smoothing parameter, $\sigma_{\rm sm}$, accounts for the instrumental resolution, $\sigma_{\rm inst}$, and the line-of-sight velocity dispersion of the star cluster, $\sigma_{\rm 1D}$. 

\subsection{Synthetic Spectra vs X-Shooter Spectra}\label{sec:comparison}
Once the resolution of the synthetic spectrum and that of the observations are similar, the ISPy3 tool matches the continuum of the X-shooter spectrum with that of the model using a cubic spline or polynomial, depending on the size of the wavelength window being analyzed.
Our software allows us to define specific weights for a given pixel in the X-Shooter spectrum, anything between 0.0 to 1.0. To avoid contamination from emission lines or other problematic features in the IL observations we have set the weights of these wavelength regions to 0.0, to exclude them from our analysis.\par

\subsection{Modelling uncertainties and systematics}
In previous publications we have assessed the various uncertainties and systematics introduced by the different choices made in the modelling. As mentioned in Section \ref{sec:stel_par}, \citet{her17} showed that the differences in the inferred abundances applying method [b] and [c] are very minor, of the order of $\lesssim$0.1 dex. Additionally, studying the integrated light of a large sample of stellar clusters \citet{her18} found overall differences $<$0.1 dex when comparing abundances inferred using solar-scaled (PARSEC) and $\alpha$-enhanced isochrones (Dartmouth, \citealt{dot07}). Regarding the uncertainties introduced by the different IMF selection, \citet{her19} calculate abundances adopting a Kroupa IMF \citep{kro01} and a Salpeter IMF \citep{sal55} and found that the uncertainties introduced by this selection is on average of the order of $<<$0.1 dex. Overall, it has been found that the systematic uncertainties should be of the order of those quoted in Table \ref{table:wabun} or smaller.

\section{Results}\label{sec:results}
Before initiating our abundance analysis described in Sections \ref{sec:stel_par}-\ref{sec:comparison}, we first obtain an estimate of the radial velocities ($v_{\rm rv}$) of the individual YMCs. We list in the second column of Table \ref{table:derived} our inferred $v_{\rm rv}$ and their uncertainties. We note that the inferred radial velocities for the different YMCs agree with the velocity maps for NGC 1313 by \citet{ryd95}.\par
As part of the abundance analysis, the ISPy3 software repeats the steps described in Section \ref{sec:ana}, modifying the chemical abundances in each iteration. We fit for the individual abundances, converging on the best value by minimizing $\chi^{2}$ and applying a golden-section search technique.\par
As a first step in our analysis we fit for the overall metallicity, [Z], and the smoothing parameter, $\sigma_{\rm sm}$, simultaneously. This is done by scanning the X-Shooter UVB and VIS wavelength ranges, 200\AA$\:$ at a time. For each YMC we select the initial isochrone adopting the ages in Table \ref{table:ymc_prop} and a metallicity of [Z] = $-$0.3 dex \citep{wal97}. We find that in general using the stellar parameters from the initial isochrone we successfully converge to an overall metallicity of [Z] $\sim-$0.3 ($\pm$0.1) dex for each of the clusters. These overall metallicities are solely used as a scaling factor in the following stages where we measure the individual element abundances. \par
To measure the abundances of the elements of interest we adopt the predefined wavelength windows used in \citet{her17}, as these have shown to help minimize the degree of blending. We begin by fitting those elements with the largest number of lines in the X-Shooter UVB and VIS wavelength ranges, one at a time and setting the smoothing parameter fixed. We first measure Fe, followed by Ti, Ca, Mg, Ni, and Cr. We attempted to measure the abundances of Sc and Mn, however, the ISPy3 software was not able to converge on a final value. We list the individual elements, wavelength bins, best-fit abundances and their corresponding 1$\sigma$ uncertainties calculated from the $\chi^2$ fit,  in Tables \ref{table:derived_201} through \ref{table:derived_505} in the Appendix section. In our previous studies we have found that the standard deviation, $\sigma_{\rm STD}$, is more representative of the actual measurement uncertainties, given that the scatter in the individual measurements is larger than the formal errors based on the $\chi^{2}$ analysis. Similar to the work of \citet{her17}, we convert the measured $\sigma_{\rm STD}$ into errors on the mean abundances adopting their equation (5). In Table \ref{table:wabun} we list the weighted average abundances and their corresponding uncertainties ($\sigma_{\rm STD}$). Lastly, we show example synthesis fits for each of the targets studied here in Appendix Figures \ref{fig:201_mod}-\ref{fig:505_mod2}. 

\subsection{Balmer emission}
\begin{figure}
   	  \centerline{\includegraphics[scale=0.6]{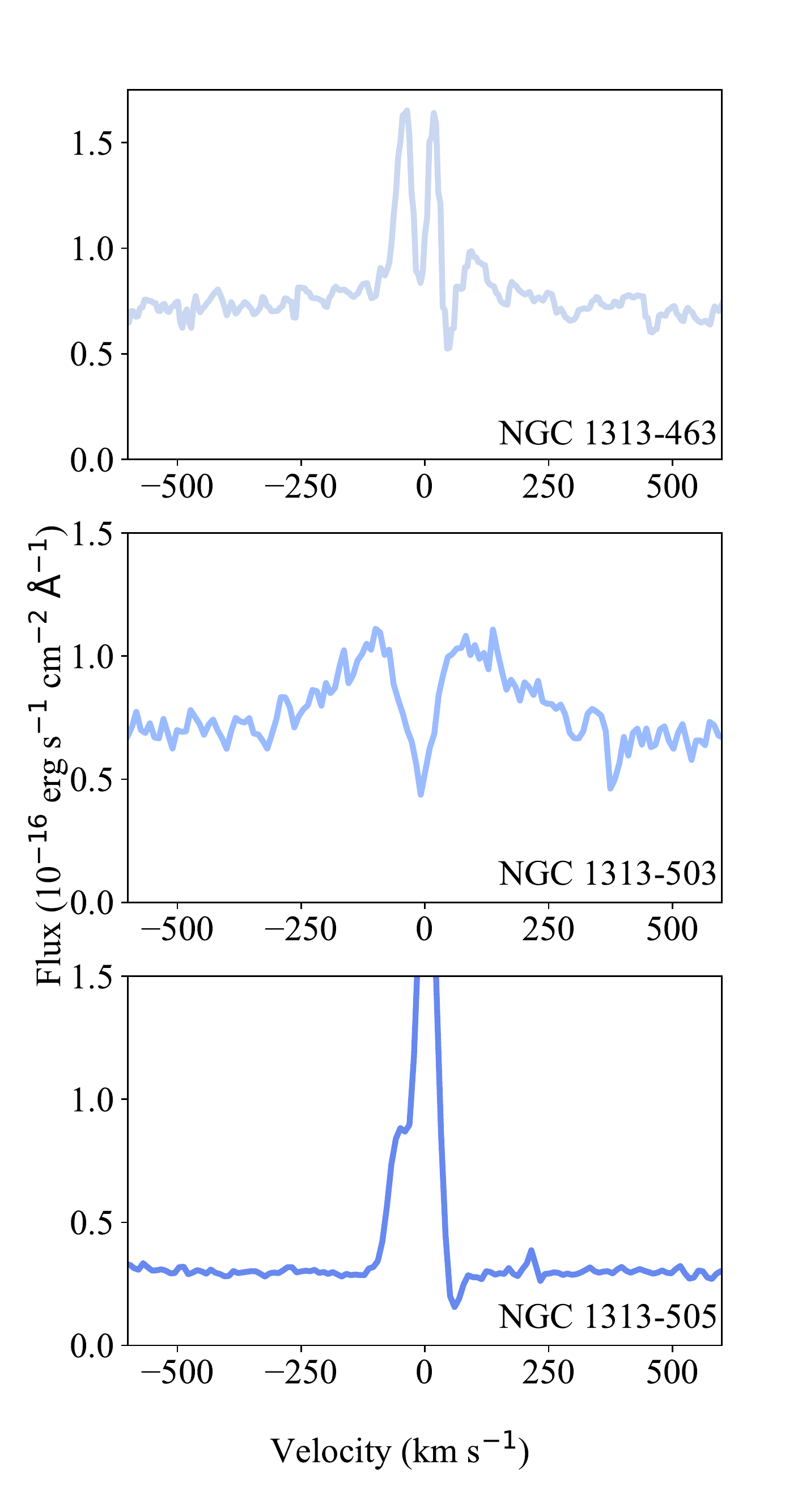}}
      \caption{H$\alpha$ emission in a subsample of our NGC 1313 YMCs. The velocities are given in the reference frame of the YMCs.}
         \label{fig:balmer}
            \end{figure}

We note that three of the YMCs analyzed here displayed strong emission lines, particularly in H$\alpha$. The three different clusters, NGC 1313-463, NGC 1313-503, and NGC 1313-505, exhibit slightly different emission profiles (Figure \ref{fig:balmer}). The H$\alpha$ profile observed in YMC NGC 1313-503 is comparable to those identified in \citet{her17} for YMCs NGC 1313-379 and NGC 1705-1. The nature of the broad profiles, similar to those observed in these three clusters, is briefly discussed by \citet{mel85}, where they proposed that one possibility for the origin of this emission in H$\alpha$, other than ionized gas, is the presence of Be stars. These objects are identified as B-type non-supergiant stars with rotational velocities of several hundred km s$^{-1}$ which display strong and broad H$\alpha$ emission features \citep{mar82,tow04}. Additionally, studies have found an enhancement of Be stars in stellar clusters with ages $<$100 Myr \citep[e.g.,][]{mat08}. The broad H$\alpha$ emission observed in NGC 1313-503, reaching velocities of $\sim$300 km s$^{-1}$, along with its estimated age of 20 Myr, hint at the presence of Be stars. \par
The H$\alpha$ profile observed in NGC 1313-463, a cluster with an estimated age of 40 Myr, appears to be slightly more complex. From visual inspection we believe the emission could be originating from both a population of Be stars (creating a slightly broad component, reaching velocities of $\lesssim$200 km s$^{-1}$), and from ionized gas surrounding the cluster (much narrower component). Lastly, the emission features observed in NGC 1313-505 appear to be originating primarily from gas, although a population of Be stars cannot be fully discarded. We highlight that any wavelength regions contaminated by emission features, irrespective of their origin, are masked in our abundance analysis. \par

\begin{table}
\caption{Derived properties of YMCs in NGC 1313}
\label{table:derived}
\centering 
\begin{tabular}{lcc}
\hline \hline \hline
Cluster & $v_{\rm rv}$ & $\sigma_{\rm 1D}$ \\
& (km s$^{-1}$) & (km s$^{-1}$) \\
\hline\\
NGC1313-201 & 403 $\pm$ 2 & 6.8 $\pm$ 3.5\\
NGC1313-439 & 483 $\pm$ 4 & 5.6 $\pm$ 3.8\\
NGC1313-463 & 484 $\pm$ 2 & 5.5 $\pm$ 5.3\\
NGC1313-503 & 510 $\pm$ 7 & 5.0 $\pm$ 3.4\\
NGC1313-505 & 469 $\pm$ 2 & 6.4 $\pm$ 1.5\\

\hline
\end{tabular}
\end{table}

\begin{table*}
\caption{Derived abundances.}
\label{table:wabun}
\centering 
\begin{tabular}{lcccccc}
\hline \hline \hline
Cluster & [Fe/H] & [Ca/Fe] & [Ti/Fe] & [Mg/Fe] & [Cr/Fe] & [Ni/Fe] \\
& (dex) & (dex) & (dex) & (dex) & (dex) & (dex)  \\
\hline\\
NGC1313-201 & $-$0.68 $\pm$ 0.15 & $+$0.10 $\pm$ 0.17 & $-$0.18 $\pm$ 0.38 & $+$0.58 $\pm$ 0.32 & $+$0.07 $\pm$ 0.20 & $-$0.04 $\pm$ 0.27\\
NGC1313-439 & $-$0.26 $\pm$ 0.16 & $-$0.09 $\pm$ 0.18 & $+$0.31 $\pm$ 0.48 & $-$0.17 $\pm$ 0.34 & $+$0.22 $\pm$ 0.64 & $-$0.05 $\pm$ 0.23\\
NGC1313-463 & $-$0.37 $\pm$ 0.10 & $+$0.06 $\pm$ 0.37 & $-$0.40 $\pm$ 0.17 & $-$0.63 $\pm$ 0.19 & $+$0.27 $\pm$ 0.13 & $+$0.09 $\pm$ 0.21\\
NGC1313-503 & $-$0.03 $\pm$ 0.08 & $-$0.20 $\pm$ 0.21 & $-$0.22 $\pm$ 0.19 & $+$0.03 $\pm$ 0.24 & $-$0.21 $\pm$ 0.08 & $+$0.51 $\pm$ 0.32\\
NGC1313-505 & $-$0.13 $\pm$ 0.08 & $-$0.06 $\pm$ 0.21 & $-$0.25 $\pm$ 0.14 & $+$0.70 $\pm$ 0.70 & $-$0.15 $\pm$ 0.14 & $-$0.24 $\pm$ 0.08\\

\hline
\end{tabular}
\end{table*}

 \section{Discussion}\label{sec:discussion}
 In this section we discuss our abundance measurements from both $\alpha$- and Fe-peak elements, and their implications on the enrichment history of NGC 1313. We also discuss in detail the abundance trends observed in this nearby galaxy and compare them to those studied in different environments, such as the MW and the LMC.
 
  \subsection{Individual abundances}
  \subsubsection{$\alpha$-elements}
\begin{figure}
   
   	  \centerline{\includegraphics[scale=0.55]{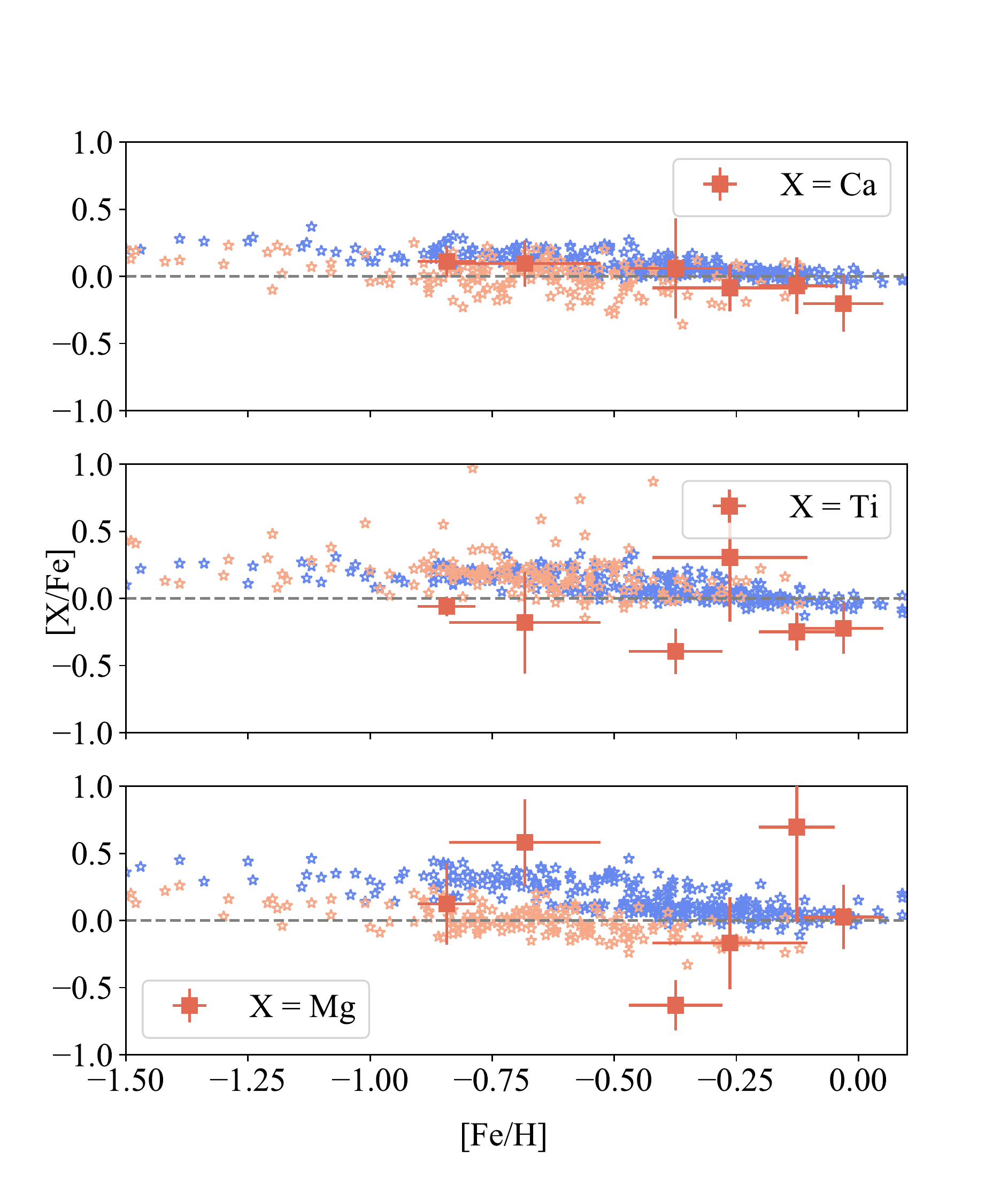}}
      \caption{[$\alpha$/Fe] ratios as a function of [Fe/H] abundances for three different elements shown in the legend. The square symbols show the abundances inferred as part of this work. Blue stars show the MW abundances by \citet{red03,red06}. Orange stars show the LMC abundances by \citet{vander13}. }
         \label{fig:ngc1313_abun_alpha}
            \end{figure}

Known as $\alpha$ elements, Ca, Ti and Mg, are primarily produced by high-mass stars and ejected to the interstellar medium through core-collapse supernovae \citep{woo95}. Due to their nature, $\alpha$-elements are typically used to gain insight into the period of time when core-collapse supernovae shaped the chemical evolution of the host galaxy. Since Fe-peak elements are also produced in massive stars, the result is a constant [$\alpha$/Fe] ratio during those epochs dominated by core-collapse supernovae. Typically, the [$\alpha$/Fe] ratios begin to decline when type Ia supernovae (SNIa) start to drive the chemical enrichment, as these release large amounts of Fe-peak elements \citep{tim03}. Overall, these are the general trends we observed in the MW. Additionally, Galactic abundances of $\alpha$ elements appear to correlate with each other. Such a trend can be clearly observed in Figure \ref{fig:ngc1313_abun_alpha}, where we show in blue stars the abundances of field stars in the MW as measured by \citet{red03,red06}. We also show in Figure \ref{fig:ngc1313_abun_alpha} as orange stars the observed abundances in the LMC, a high-mass dwarf galaxy, published by \citet{vander13}. We note that the LMC and NGC 1313 share a few properties. Both galaxies are classified as spiral galaxies, the LMC as SBm and NGC 1313 as SB(s) \citep{san81}, with comparable metallicities \citep{wal97, rus92}.
\\
The $\alpha$-element abundances presented here, are the first-ever measured in stellar populations in NGC 1313. The abundances for all the clusters in our sample, including those from \citet{her17}, are plotted as orange squares in Figure \ref{fig:ngc1313_abun_alpha} as a function of [Fe/H]. The [Ca/Fe] ratios for all of the clusters in NGC 1313 tend to decrease with increasing [Fe/H], following a trend comparable to that observed in the LMC, and with slightly more depleted Ca abundances than those from the MW field stars.\\
The [Ti/Fe] values observed in the YMC sample show a much higher scatter than that of Ca, however, a similarly high scatter is seen in the field stars of the LMC. This larger scatter in the LMC [Ti/Fe] ratios is primarily introduced by the stars in the inner disk, compared to the abundances in the LMC bar \citep{vander13}. It is also clear from Figure \ref{fig:ngc1313_abun_alpha} that overall, the Ti abundances are lower in NGC 1313, compared to those in the LMC for a given metallicity. \\
We observe a comparable scatter in the [Mg/Fe] ratios in NGC 1313, to  that in the [Ti/Fe] abundances. However, we also highlight that the uncertainties in the Mg abundances are much higher than those from Ti, particularly for NGC 1313-505. For this cluster we infer [Mg/Fe] = $+$0.70 $\pm$ 0.7 dex, with abundances compatible to those in the field in the MW and LMC, within the uncertainties. \\

\subsubsection{Fe-peak elements}
\begin{figure}
   
   	  \centerline{\includegraphics[scale=0.55]{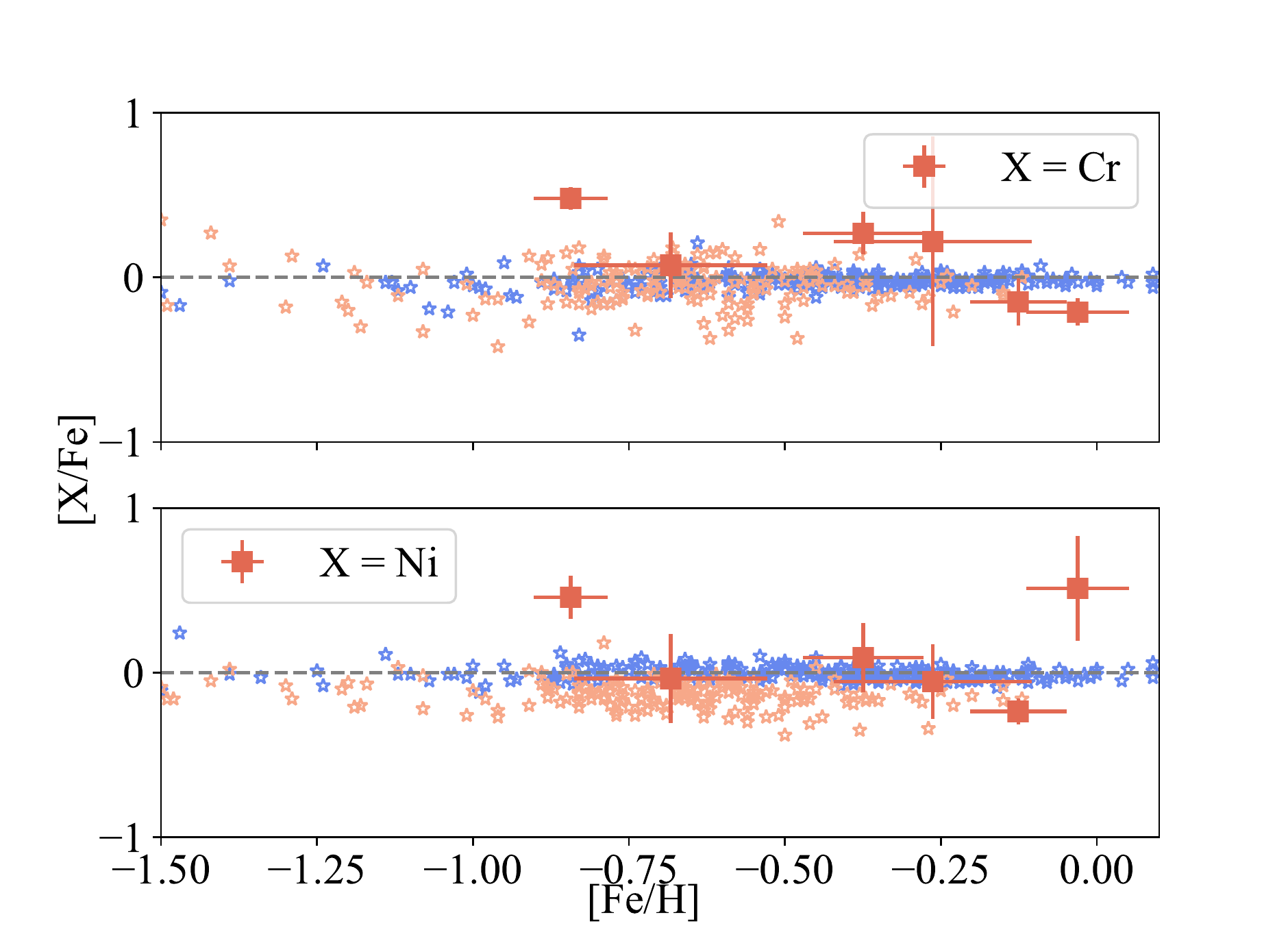}}
      \caption{Same as in Figure \ref{fig:ngc1313_abun_alpha} for Fe-peak elements.}
         \label{fig:ngc1313_abun_Fepeak}
            \end{figure}

In Figure \ref{fig:ngc1313_abun_Fepeak} we show the NGC 1313 YMC abundances for the Fe-peak elements Cr and Ni, and compared them with those observed in stars in the MW \citep{red03,red06} and the LMC \citep{vander13}. Both of these Fe-peak elements are primarily produced by Type Ia SNe. However, in spite of sharing a common origin, they exhibit slightly different patterns.\par
Cr abundances, both in the LMC and the MW, display a much higher scatter than that observed in Ni. The stellar abundances in these two environments are typically flat around [Cr/Fe]= 0.0 dex. The [Cr/Fe] ratios measured for our sample of YMCs in NGC 1313 appear to follow a comparable trend to those observed in the MW and LMC. The only cluster that appears to slightly deviate from this trend is the most metal-poor YMCs, NGC 1313-379, studied in \citet{her17}, displaying a subtle Cr enhancement. \par
Ni also exhibits a flat distribution in both the MW and the LMC. The trend observed in the LMC, however, is consistently sub-solar. Overall, the Ni abundances measured in NGC 1313 follow the flat trends of the MW and LMC, again with the exception of the most-metal poor YMCs, as well as the most-metal rich target. \par

 \subsection{Stellar Abundance gradients}            
 Over the last couple of decades, studies investigating the chemical variations as a function of galactocentric distances in spiral galaxies have provided essential constraints on the evolution of these systems. The majority of these studies, greatly relying on oxygen abundances of H II regions, hint at an inside-out growth of galactic discs \citep{pil14, bre15, ho15, lia18}. In this section we present our results on the stellar abundance gradients for Fe, and the $\alpha$ elements measured in this work. \par
\begin{figure*}
   
   	  \centerline{\includegraphics[scale=0.65]{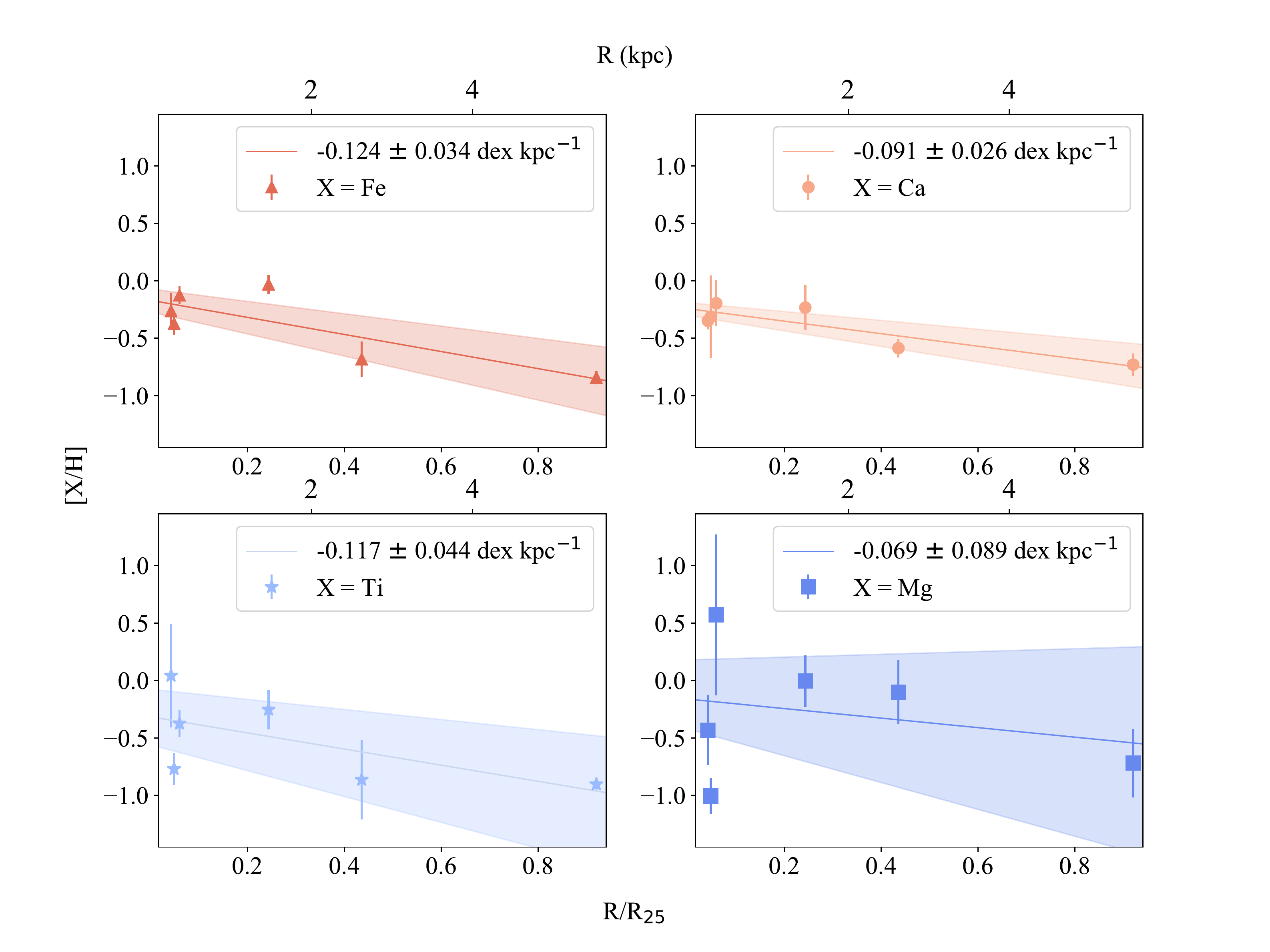}}
      \caption{Abundances as a function of galactocentric distance normalized to isophotal radius. In each panel we display with solid lines the linear regressions for each set. We show in the legend of each panel the inferred abundance gradient, as well as the corresponding element. }
         \label{fig:ngc1313_abun_grad}
   \end{figure*}

 In Figure \ref{fig:ngc1313_abun_grad} we show in different panels the abundances as a function galactocentric distance for the individual elements, Fe, Ca, Ti, and Mg. We compute linear regressions for each element and show the inferred slopes with a solid line. Given that  the theorems underlying least squares regressions assume asymptotic normality (meaning large-N samples), a condition that does not apply to our limited sample, we estimate our abundance gradients and their uncertainties applying a linear regression with non-parametric bootstrapping, where we apply the bootstrapping on the residuals, not the fitted parameters themselves. Briefly summarized, we find the optimal linear regression on our original dataset, extract the residuals from the best fit, generate new equal-size samples using the residuals and fitting linear regressions on the new samples. This resampling method allows for a more accurate estimate of the uncertainties in our inferred parameters. \par
 We measure an Fe gradient of $-$0.124 $\pm$ 0.034 dex kpc$^{-1}$. We find a slightly shallower gradient for Ca, $-$0.091 $\pm$ 0.026 dex kpc$^{-1}$. And finally, the gradients for Ti and Mg show a much larger scatter, with values of $-$0.117 $\pm$ 0.044 dex kpc$^{-1}$ and $-$0.069 $\pm$ 0.089 dex kpc$^{-1}$, respectively. We estimate a weighted average $\alpha$-abundance gradient of $-$0.096 $\pm$ 0.009 dex kpc$^{-1}$, which is only slightly flatter than that observed for Fe. It is important to highlight this subtle difference as it is expected that these elements, Fe-peak and $\alpha$, with different nucleosynthetic origins, imprint a different signature in the observed gradients. For example, in the past such a behaviour has been predicted and observed in the MW \citep{chi01, pal20}. Overall, negative abundance gradients are obtained when the star formation has been more efficient in the inner regions of the galaxy, compared to those in the outer regions of the disk, which based on our study, appears to be the case in NGC 1313.  \par
Lastly, we perform a final test as the abundance gradients in Figure \ref{fig:ngc1313_abun_grad} appear to be primarily driven by the individual point at R/R$_{25}$$\sim$0.9. We remove this individual point and estimate new gradients with the remaining points at R/R$_{25}<$0.8. Overall we find that the abundance gradients for most elements remain within the uncertainties quoted in Figure \ref{fig:ngc1313_abun_grad}, with the exception of Mg. The gradient for this particular element becomes slightly positive with much larger uncertainties, strongly highlighting the effects of the large scatter on the inferred Mg gradient.

  \subsection{Comparison to ionized-gas abundances}
 Past studies have investigated the distribution of metals in NGC 1313. \citet{pag80} analyzed the observations of six \ion{H}{2} regions and found a uniform gas-phase abundance distribution, with no clear gradient. Similarly, through the spectroscopic analysis of a much larger sample of 33 \ion{H}{2} regions, \citet{wal97} reported a flat gas-phase abundance distribution across the disk. These results made NGC 1313 the most massive barred galaxy to lack a radial abundance gradient in the ionized-gas
component. We highlight that these abundance studies relied on strong-line calibrations, which are based on the flux ratios of some of the strongest forbidden lines. It is well known that different strong-line calibrations return different abundance gradients with systematic offsets in the measured abundances \citep{kew08, mai19}, particularly in high-metallicity environments \citep{bre16}. Gas-phase metallicities estimated using the $T_e$-based method are considered to be more reliable than those inferred from strong-line calibrations \citep{mai19}.\par
\begin{figure}
   
   \centerline{\includegraphics[scale=0.53]{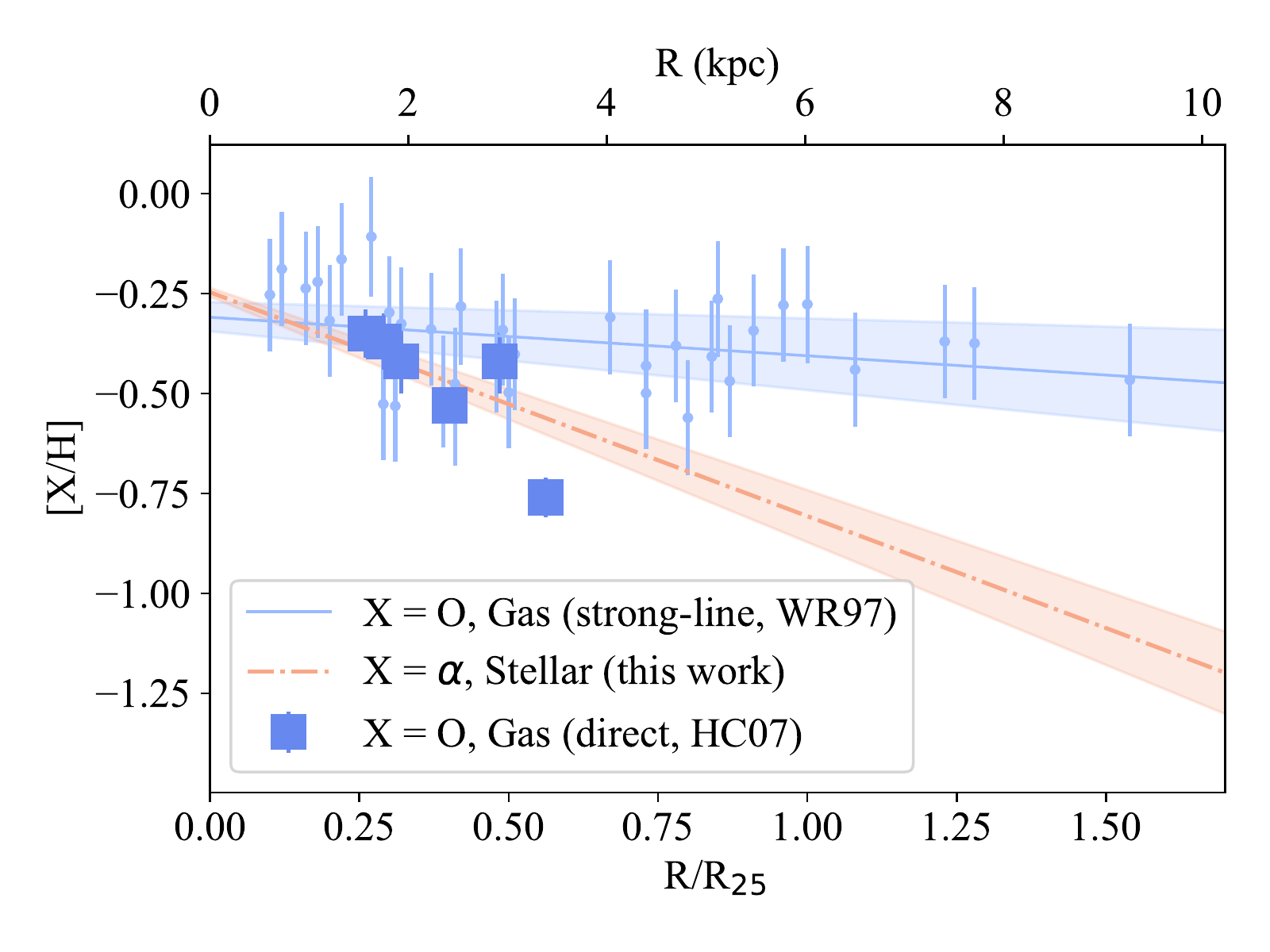}}
      \caption{Abundances as a function of galactocentric distances. We show with a blue solid line the linear regression for the gas-phase oxygen abundances inferred with the strong-line O3N2 calibration by \citet{pet04}, using the fluxes by \citet{wal97}. We show with a dashed-dotted orange line our stellar weighted average $\alpha$-abundance gradient. The blue squares display the depletion-corrected direct oxygen abundances by \citet{had07}.} 
         \label{fig:ngc1313_neb_grad}
   \end{figure}

 In Figure \ref{fig:ngc1313_neb_grad} we compare our stellar weighted average $\alpha$-abundance gradient to the much shallower gradient inferred from the \ion{H}{2} regions studied in \citet[][herein after WR97]{wal97}, $-$0.016 $\pm$ 0.008 dex kpc$^{-1}$. The nebular abundances were estimated using the strong-line [\ion{O}{3}]/[\ion{N}{2}] calibration by \citet{pet04} along with the measured fluxes by WR97. From this figure, it is clear that the overall trends imprinted in the stellar and gas-phase abundances are remarkably different. A puzzling effect as both of these components are expected to describe the present-day metallictiy distribution of NGC 1313. Studies such as that by \citet{bre16} and \citet{her21} have previously shown that strong-line indicators produce nebular abundance gradients that are significantly shallower than those inferred from young stellar populations.\par
We expand our nebular and stellar abundance gradient comparison to include the direct abundances for six \ion{H}{2} regions studied in \citet{had07}. In order to perform a meaningful comparison between the direct oxygen abundances and those measured from our YMC sample, we must correct the oxygen gas-phase metallicities for depletion onto interstellar dust grains. Overall, studies estimate depletion factors for oxygen that are between $-$0.08 and $-$0.2 dex \citep{mes09, jen09, pei10, sim11}. Similar to the work of \citet{bre16}, for simplicity we adopt a depletion correction factor of $-$0.1 dex. In Figure \ref{fig:ngc1313_neb_grad} we show with square symbols the depletion-corrected [O/H] abundances by \citet{had07}. We note that due to the limited size of their \ion{H}{2}-region sample, \citet{had07} report no evidence for a radial dependence in their inferred abundances. Figure \ref{fig:ngc1313_neb_grad}, however, hints at a possibly steep gradient for the ionized gas primarily driven by the single abundance measurement at R $\sim$ 3 kpc. This figure clearly shows that the direct oxygen abundances appear to be much better aligned to the stellar abundance gradient, than to the strong-line abundance trend by WR97. To further confirm the presence of such a steep gradient in the gas-phase metallicities in NGC 1313, the sample of \ion{H}{2} regions with available [\ion{O}{2}] $\lambda$3727, [\ion{O}{3}] $\lambda$4363 and [\ion{O}{3}] $\lambda$5007 nebular emission features (to obtain direct abundances) needs to be expanded to cover galactocentric distances close to the nucleus, as well as in the outskirts of the galaxy. Given the limitations of the strong-line calibrations, it is critical that abundance studies comparing the gas-phase and stellar populations adopt whenever possible $T_e$-based nebular calibrations, otherwise, any information on their chemical evolution or mixing timescales might be misconstrued. 
 
  \subsection{Age-metallicity relation in NGC 1313}
 Age-metallicity relations are important for understanding the chemical evolution of individual galaxies. These types of relations are able to shed some light on correlations between the age of the stellar clusters and possible interactions or close encounters with other systems. \citet{liv13} studied the age-metallicity relation in the LMC by analyzing a sample of small open clusters. They report a metallicity gradient with higher abundances for the younger clusters. Such a trend is of course expected in the normal evolution of stellar populations in galaxies. \citet{liv13} also identify a considerable increase in metallicity around the ages of $\sim$600 Myr, which they propose could be the result of the most recent encounter between the LMC and the SMC. \par
\begin{figure}
   \centerline{\includegraphics[scale=0.57]{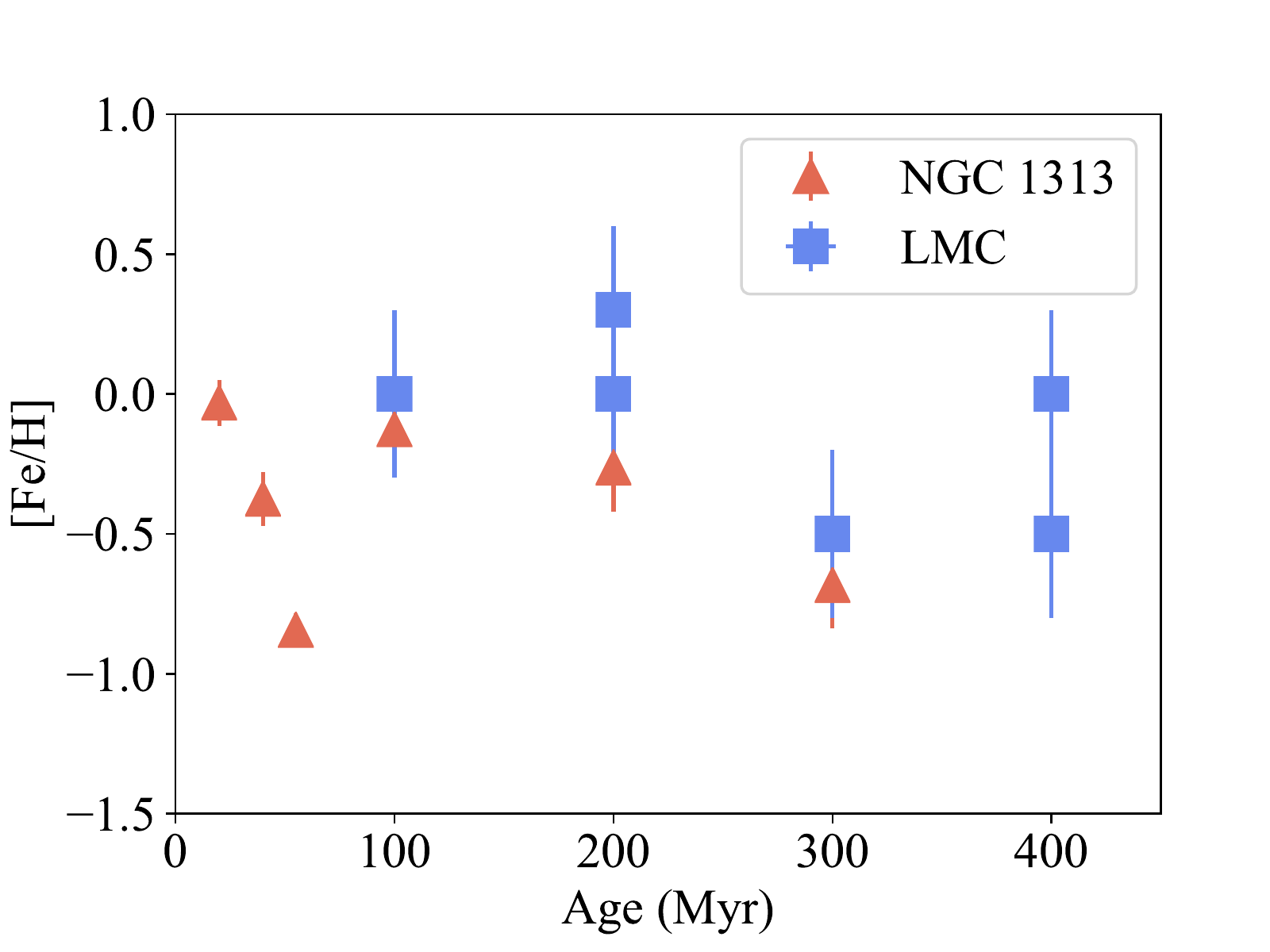}}
      \caption{Age-metallicity trends. We show with orange triangles the values of the YMCs in NGC 1313. The blue squares display the metallicities and ages for a sample of open clusters in the LMC by \citet{liv13}.} 
         \label{fig:age_Z}
   \end{figure}

 Our sample of YMCs in NGC 1313 cover ages between 20 and 300 Myr. In Figure \ref{fig:age_Z} we display the age-metallicity trend for NGC 1313 with orange triangles, and compare it to the relation observed in the LMC (shown as blue squares) for clusters with comparable ages. Overall, there is a clear trend where clusters in NGC 1313 with ages $\geq$100 Myr began to show a steady decline in their metallicities. One YMC appears to deviate from this subtle trend, NGC 1313-379, with an age of 55 Myr and a metallicity of [Fe/H] = $-$0.84 $\pm$ 0.07 dex. Interestingly, through the study of resolved stellar populations \citet{sil12} find that the star formation histories suggest a burst in the south-western region, right at the location of NGC 1313-379. This in contrast to the rest of the galaxy, the bar and arms, which exhibits a constant star formation history. Their findings support a scenario where the south-west region of NGC 1313 is experiencing an interaction with a relatively small companion satellite.  Previous to the work by \citet{sil12}, several other studies had found evidence of satellite interactions in this same location \citep{bla81, pet94, lar07}. The age-metallicity relation observed as part of this work (Figure \ref{fig:age_Z}) supports a scenario of constant star formation history throughout the galaxy, with a possible recent burst in star formation close to the region where NGC 1313-379 is located. 
 
\section{Conclusions}\label{sec:con}
We carry out the most extensive detailed abundance analysis of a sample of YMCs in the barred spiral galaxy NGC 1313 to date. Our analysis relied on X-Shooter/VLT spectroscopic observations and the method and code developed by \citet{lar12}. The main results of our work are summarized as follows: 
\begin{itemize}
\item We obtain the first stellar abundance gradients for NGC 1313. We infer an Fe gradient of $-$0.124 $\pm$ 0.034 dex kpc$^{-1}$ and a weighted average $\alpha$ gradient of $-$0.096 $\pm$ 0.009 dex kpc$^{-1}$. We highlight the importance of this subtle difference between the two values as Fe-peak and $\alpha$ elements have different nucleosynthetic origins. 
\item We compare our stellar abundance gradient to those in the literature for the gas-phase component. Past studies of \ion{H}{2} regions have reported an absence of a metallicity  gradient in NGC 1313. Comparing our stellar abundances with those inferred for the ionized gas using the $T_e$-based method, we find a much better agreement than those studies using strong-line calibrations. We highlight, however, that to confirm the presence of a steep gradient in the gas-phase metallicities, the sample of \ion{H}{2} regions with available [\ion{O}{2}] $\lambda$3727, [\ion{O}{3}] $\lambda$4363  and [\ion{O}{3}] $\lambda$5007 features needs to be expanded to cover the inner- and outer-most regions of the galaxy. 
\item We investigate the age-metallicity trends in NGC 1313. We observed a general trend where the metallicity declines with age. YMC NGC 1313-379 appears to slightly deviate from this trend, displaying a lower metallicity than that expected for a 55 Myr cluster. Our work supports a scenario of constant star formation throughout the galaxy, with a possible burst in the south-western region, where NGC 1313-379 is located. 
\end{itemize} 

Integrated-light abundance analysis continues to be a critical tool to study the chemical enrichment histories of galaxies outside of the Local Group. We highlight that abundance studies comparing the metallicities from different components, e.g., stellar and multi-phase gas (neutral and ionized; \citealt{her21}), are essential to understand the multiple mechanisms driving the evolution of galaxies.\par 
\appendix
\twocolumngrid
In this section we present a complete set of modelled spectra which includes our final measured abundances. In Tables \ref{table:derived_201} to  \ref{table:derived_505} we list the derived abundances for the different wavelength bins. 
\begin{table}
\caption{Chemical Abundances for NGC1313-201}
\label{table:derived_201}
\centering 
\begin{tabular}{lccc}
\hline \hline 
Element & Wavelength[\r{A}] & Abundance & Error \\
\hline 
{[Fe/H]} & 4700-4800 & $-$0.854 & 0.189 \\
 & 4900-5000 & $-$0.796  & 0.091 \\
 & 5000-5100 &  $+$0.656  & 0.121\\
 & 6100-6300 & $-$0.854 & 0.180\\
 & 6300-6340 & $-$0.515  & 0.290\\
 & 8500-8700 & $-$0.725 & 0.171\\
 & 8700-8850 & $+$0.301 & 0.215\\
 {[Mg/H]} & 5150-5200 & $-$0.025 & 0.081 \\
 & 8777-8832 & $-$0.585 & 0.201\\
 {[Ca/H]} & 4445-4465 & $-$0.375 & 0.391\\
 & 6100-6128 & $-$0.512 & 0.441\\
 & 6430-6454 & $-$0.836 & 0.632\\
 & 6459-6478 & $-$ &$-$\\
 & 8480-8586 & $-$0.535 & 0.05\\
 & 8623-8697 & $-$0.695 & 0.07\\
 {[Ti/H]} & 4650-4718 & $-$1.325 & 0.311\\
 & 4980-5045 & $-$0.634 & 0.219\\
 & 6584-6780 & $-$ &$-$\\
 {[Cr/H]} & 4580-4640 & $-$0.724 & 0.149\\
 & 4640-4675 & $-$0.315 & 0.236\\
 & 4915-4930 & $-$0.633 &  0.585\\
 {[Ni/H]} & 4700-4720 & $-$0.116 & 0.391\\
 & 4825-4840 & $-$ & $-$\\
 & 4910-4955 & $-$0.415 & 0.271\\
 & 5075-5175 & $-$1.445 & 0.280\\
 & 6100-6200 & $-$0.815 & 0.431\\
 & 7700-7800 & $-$0.556 & 0.312\\

\hline
\end{tabular}
\end{table}

\begin{table}
\caption{Chemical Abundances for NGC1313-439}
\label{table:derived_439}
\centering 
\begin{tabular}{lccc}
\hline \hline 
Element & Wavelength[\r{A}] & Abundance & Error \\
\hline 
{[Fe/H]} & 4700-4800 & $-$0.168 &0.160\\
 & 4900-5000 & $-$0.738 & 0.141\\
 & 5000-5100 & $-$0.128 & 0.110\\
 & 6100-6300 & $-$0.178 & 0.110 \\
 & 6300-6340 & $-$1.228 & 0.341\\
 & 8500-8700 & $-$0.118 & 0.101 \\
 & 8700-8850 & $-$0.413 & 0.195\\
 {[Mg/H]} & 5150-5200 & $-$0.259 & 0.132\\
 & 8777-8832 & $-$0.868 & 0.210 \\
 {[Ca/H]} & 4445-4465 & $-$0.185& 0.631 \\
 & 6100-6128 & $-$0.237 &  0.433\\
 & 6430-6454 & $-$ &$-$\\
 & 6459-6478 & $-$0.488 & 0.483\\
 & 8480-8586 & $-$0.098 & 0.033\\
 & 8623-8697 & $-$0.054 & 0.055\\
 {[Ti/H]} & 4650-4718 & $+$0.522 & 0.271\\
 & 4980-5045 & $-$0.382  & 0.255\\
 & 6584-6780 & $-$ & $-$\\
 {[Cr/H]} & 4580-4640 & $-$1.428 &0.351 \\
 & 4640-4675 & $+$0.512 & 0.240 \\
 & 4915-4930 & $+$0.311 & 0.472\\
 {[Ni/H]} & 4700-4720 & $-$0.717 & 0.620\\
 & 4825-4840 & $-$0.003 & 0.787\\
 & 4910-4955 & $-$0.699 & 0.361 \\
 & 5075-5175 & $-$0.578 & 0.271\\
 & 6100-6200 & $-$ & $-$\\
 & 7700-7800 & $-$0.018 & 0.206\\

\hline
\end{tabular}
\end{table}

\begin{table}
\caption{Chemical Abundances for NGC1313-463}
\label{table:derived_463}
\centering 
\begin{tabular}{lccc}
\hline \hline 
Element & Wavelength[\r{A}] & Abundance & Error \\
\hline 
{[Fe/H]} & 4700-4800& $-$0.968 & 0.251\\
 & 4900-5000& $-$0.589 & 0.105\\
 & 5000-5100& $-$0.529 & 0.115\\
 & 6100-6300& $-$0.550 & 0.161\\
 & 6300-6340& $-$0.629 & 0.245\\
 & 8500-8700& $-$0.248 & 0.059 \\
 & 8700-8850& $-$0.209 & 0.104\\
 {[Mg/H]} & 5150-5200& $-$1.207 & 0.145\\
 & 8777-8832& $-$0.888 & 0.111\\
 {[Ca/H]} & 4445-4465& $+$0.483 & 0.259 \\
 & 6100-6128& $-$0.069 & 0.164\\
 & 6430-6454& $+$0.107 & 0.223\\
 & 6459-6478& $-$2.059 &  0.532\\
 & 8480-8586& $-$0.302 & 0.019\\
 & 8623-8697& $-$0.376 &0.033\\
 {[Ti/H]} & 4650-4718& $-$0.559 & 0.327 \\
 & 4980-5045& $-$0.839 & 0.188\\
 & 6584-6780& $-$ & $-$\\
 {[Cr/H]} & 4580-4640& $-$0.209& 0.204\\
 & 4640-4675& $+$0.038 & 0.289\\
 & 4915-4930& $+$0.061 & 0.469 \\
 {[Ni/H]} & 4700-4720& $-$ & $-$\\
 & 4825-4840& $-$ & $-$\\
 & 4910-4955& $+$0.001 & 0.321 \\
 & 5075-5175& $+$0.001 & 0.211\\
 & 6100-6200& $-$ & $-$\\
 & 7700-7800& $-$0.559 & 0.174 \\

\hline
\end{tabular}
\end{table}

\begin{table}
\caption{Chemical Abundances for NGC1313-503}
\label{table:derived_503}
\centering 
\begin{tabular}{lccc}
\hline \hline 
Element & Wavelength[\r{A}] & Abundance & Error \\
\hline 
{[Fe/H]} & 4700-4800& $+$0.375 & 0.135\\
 & 4900-5000& $-$0.155 & 0.091\\
 & 5000-5100& $-$0.196 & 0.119\\
 & 6100-6300& $-$0.045 & 0.141\\
 & 6300-6340& $+$0.024 & 0.231\\
 & 8500-8700& $+$0.306 & 0.672 \\
 & 8700-8850& $+$0.026 & 0.128 \\
 {[Mg/H]} & 5150-5200& $-$0.322 &0.122  \\
 & 8777-8832& $+$0.128& 0.079\\
 {[Ca/H]} & 4445-4465&  $+$0.258& 0.399 \\
 & 6100-6128& $+$0.608 & 0.178\\
 & 6430-6454& $+$0.656 & 0.293\\
 & 6459-6478& $+$0.629 & 0.364\\
 & 8480-8586& $-$0.232 & 0.022\\
 & 8623-8697& $-$0.431 & 0.064\\
 {[Ti/H]} & 4650-4718& $-$0.582 & 0.370 \\
 & 4980-5045& $-$0.022 & 0.142\\
 & 6584-6780& $-$0.482 & 0.158\\
 {[Cr/H]} & 4580-4640& $-$0.242 & 0.236 \\
 & 4640-4675& $-$0.231 & 0.496\\
 & 4915-4930& $-$ & $-$\\
 {[Ni/H]} & 4700-4720&$-$ & $-$ \\
 & 4825-4840&$-$ & $-$ \\
 & 4910-4955& $-$ & $-$\\
 & 5075-5175& $-$0.362 & 0.334\\
 & 6100-6200& $-$0.162 & 0.453\\
 & 7700-7800& $+$0.644 & 0.124\\

\hline
\end{tabular}
\end{table}

\begin{table}
\caption{Chemical Abundances for NGC1313-505}
\label{table:derived_505}
\centering 
\begin{tabular}{lccc}
\hline \hline 
Element & Wavelength[\r{A}] & Abundance & Error \\
\hline 
{[Fe/H]} & 4700-4800& $+$0.116 & 0.083\\
 & 4900-5000& $-$0.014 & 0.022\\
 & 5000-5100& $-$0.415 & 0.051\\
 & 6100-6300& $-$0.295 & 0.051\\
 & 6300-6340& $-$0.404 & 0.105\\
 & 8500-8700& $-$0.148 & 0.034\\
 & 8700-8850& $-$0.335 & 0.067\\
 {[Mg/H]} & 5150-5200& $+$0.975 & 0.037 \\
 & 8777-8832& $-$0.425 & 0.058  \\
 {[Ca/H]} & 4445-4465& $+$0.684 & 0.191 \\
 & 6100-6128& $+$0.275 &0.105 \\
 & 6430-6454& $+$0.295 & 0.129\\
 & 6459-6478& $+$0.744 & 0.161 \\
 & 8480-8586& $-$0.194 & 0.003\\
 & 8623-8697&$-$0.494 & 0.033\\
 {[Ti/H]} & 4650-4718& $-$0.135 & 0.171\\
 & 4980-5045& $-$0.414 & 0.109\\
 & 6584-6780& $-$0.536 & 0.182\\
 {[Cr/H]} & 4580-4640& $-$0.275 & 0.121\\
 & 4640-4675& $-$ & $-$\\
 & 4915-4930& $-$ & $-$\\
 {[Ni/H]} & 4700-4720& $-$0.414 & 0.304\\
 & 4825-4840& $-$0.416 & 0.512\\
 & 4910-4955& $-$ & $-$\\
 & 5075-5175& $-$0.364 &0.122\\
 & 6100-6200& $-$ & $-$\\
 & 7700-7800& $-$0.355 &0.091\\

\hline
\end{tabular}
\end{table}

       \begin{figure*}
   \centerline{\includegraphics[scale=0.46, angle=270]{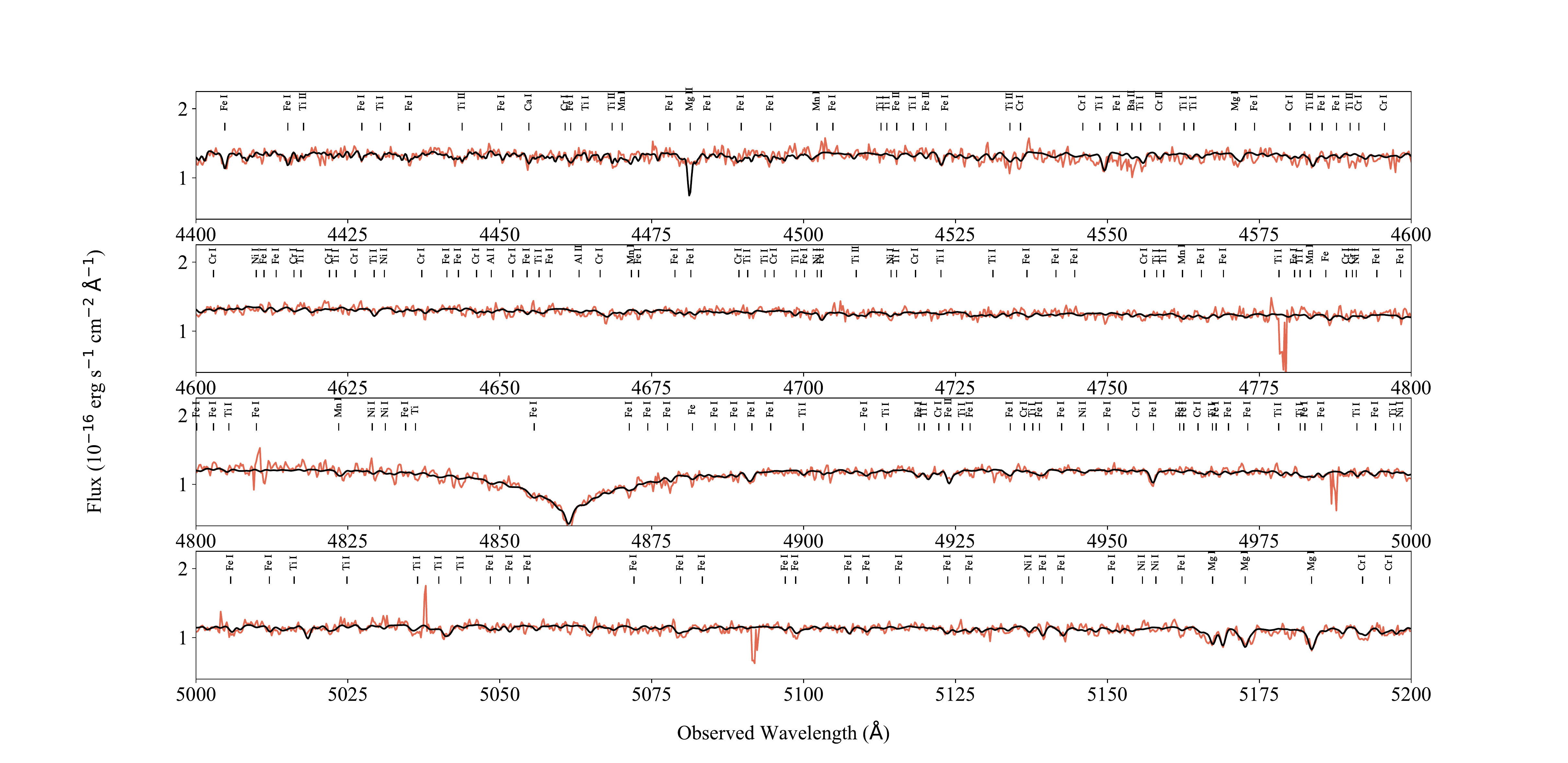}}
      \caption{Example synthesis fits for NGC 1313-201 (UVB X-Shooter coverage). In black we show the best-fitting model spectra. These models use the final abundances obtained as part of this work.}
         \label{fig:201_mod}
   \end{figure*}

       \begin{figure*}
   \centerline{\includegraphics[scale=0.46, angle=270]{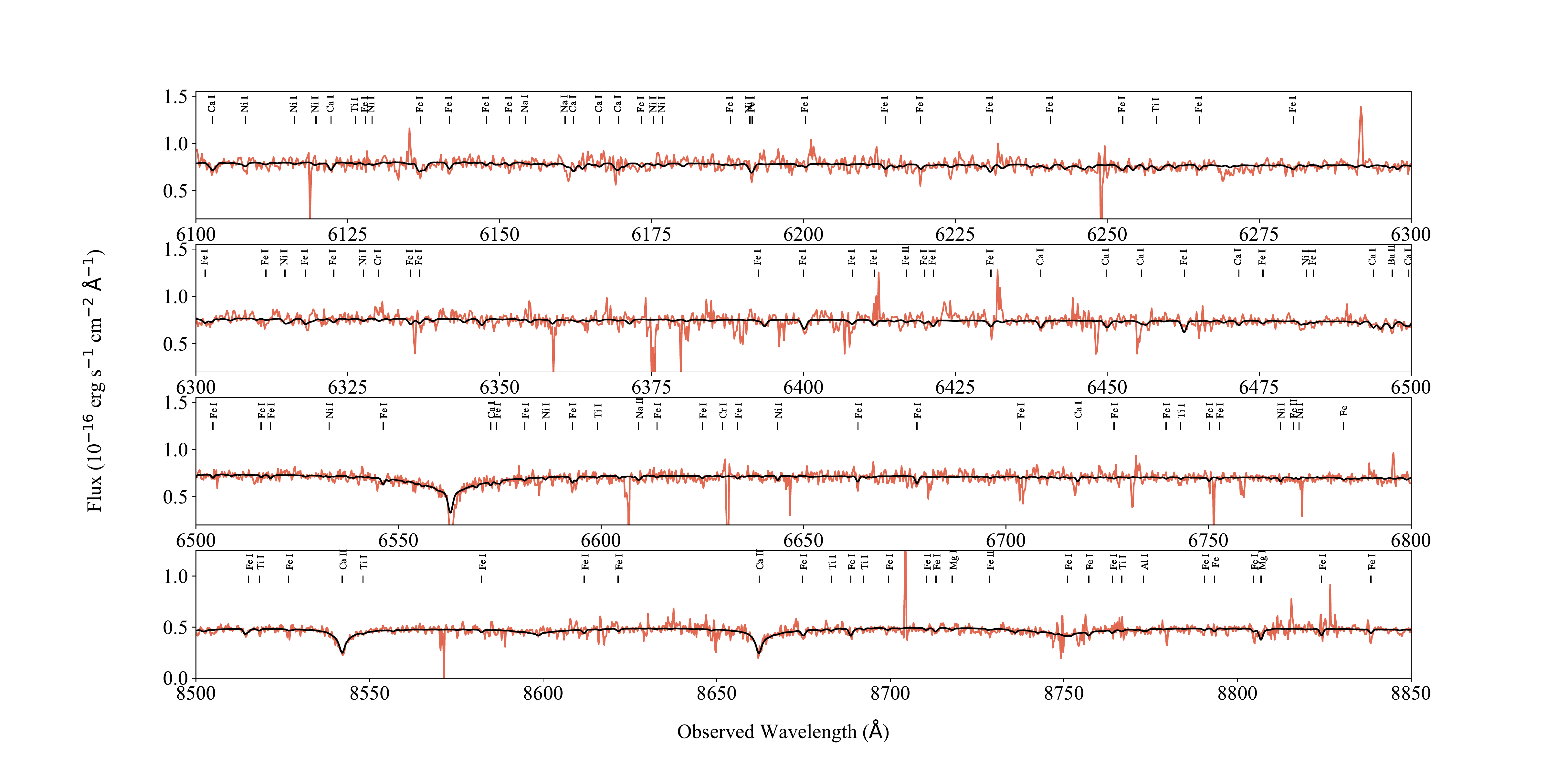}}
      \caption{Example synthesis fits for NGC 1313-201 (VIS X-Shooter coverage). In black we show the best-fitting model spectra. These models use the final abundances obtained as part of this work.}
         \label{fig:201_mod2}
   \end{figure*}

       \begin{figure*}
   \centerline{\includegraphics[scale=0.46, angle=270]{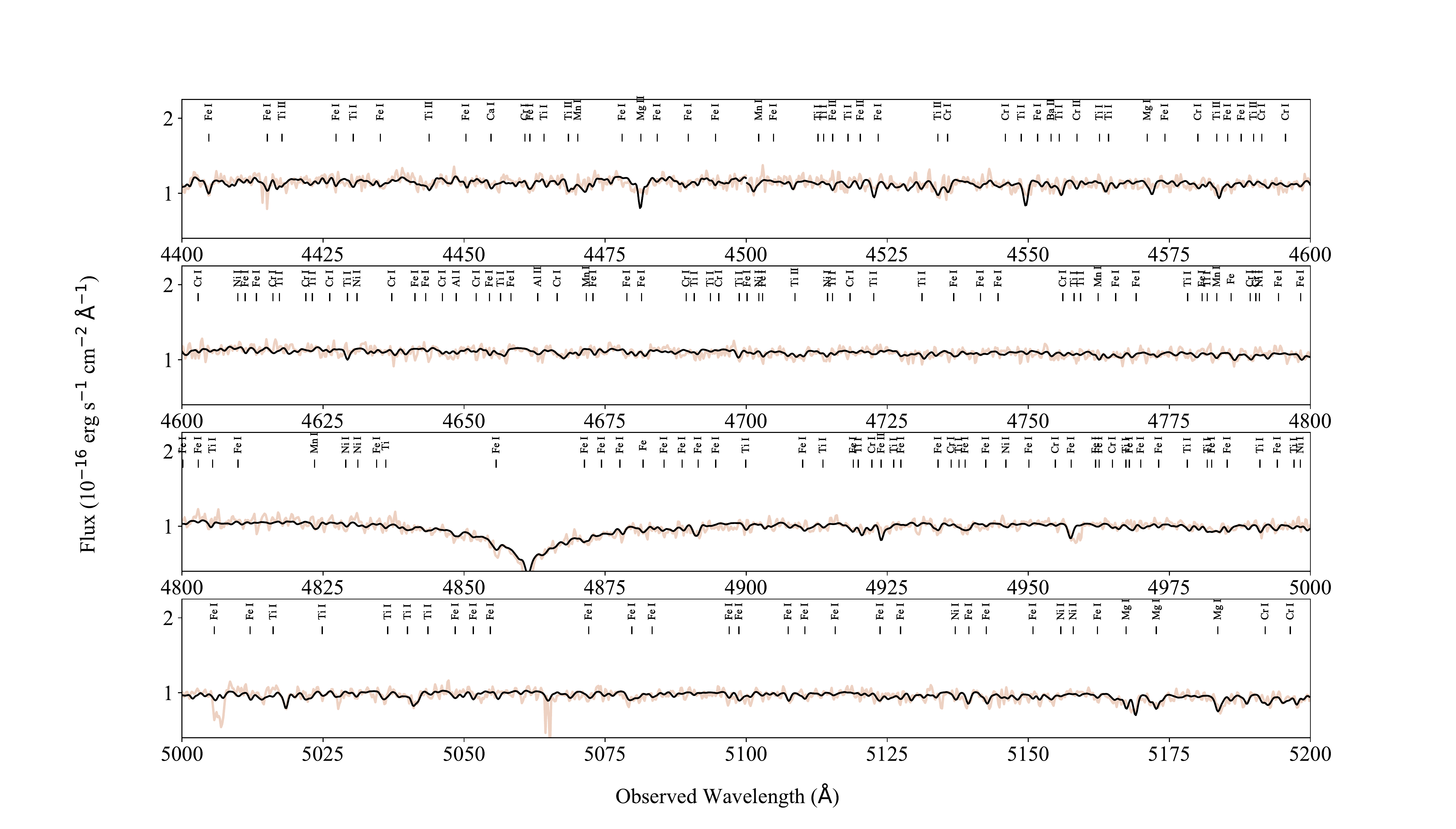}}
      \caption{Example synthesis fits for NGC 1313-439 (UVB X-Shooter coverage). In black we show the best-fitting model spectra. These models use the final abundances obtained as part of this work.}
         \label{fig:439_mod}
   \end{figure*}

       \begin{figure*}
   \centerline{\includegraphics[scale=0.46, angle=270]{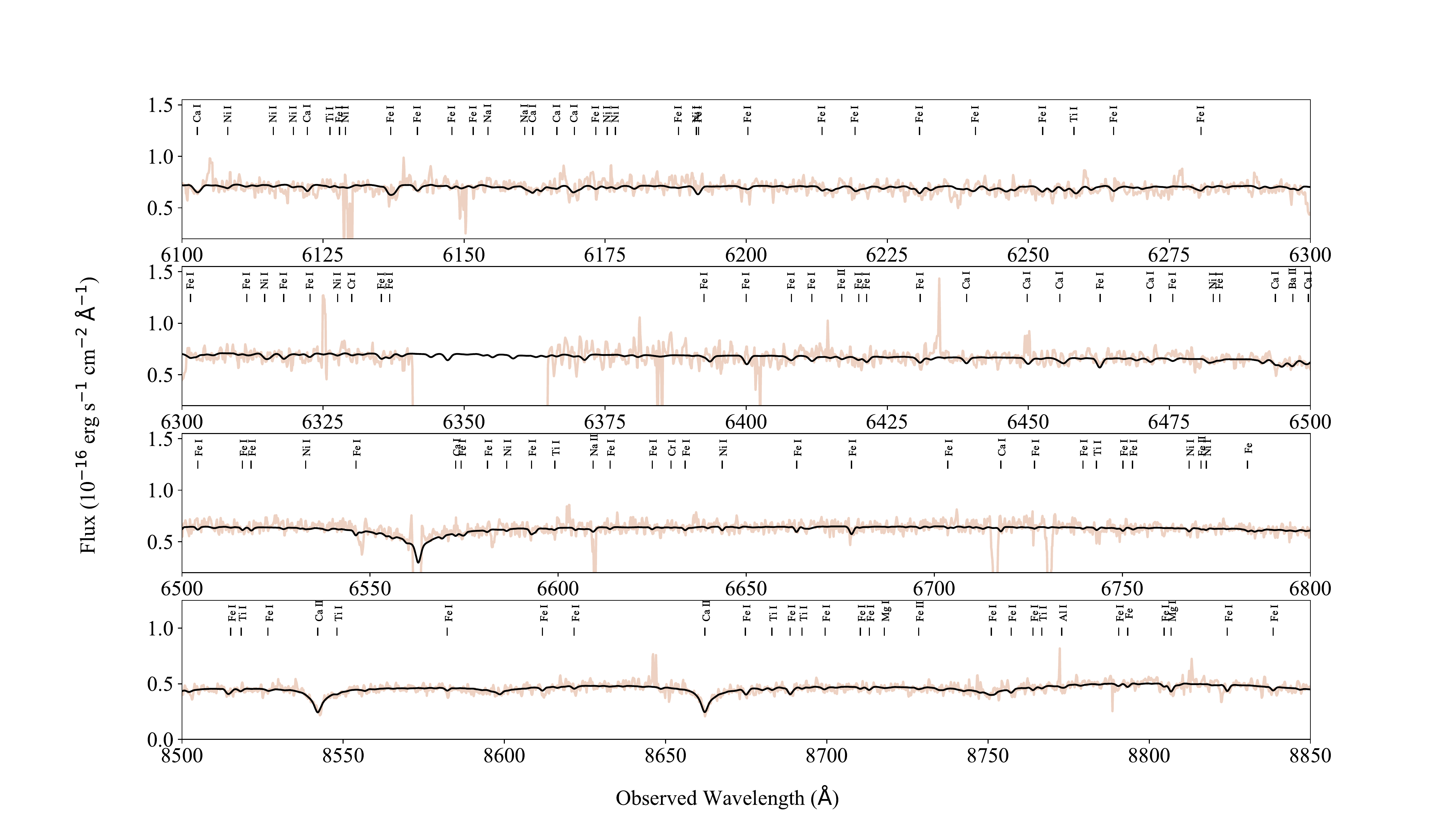}}
      \caption{Example synthesis fits for NGC 1313-439 (VIS X-Shooter coverage). In black we show the best-fitting model spectra. These models use the final abundances obtained as part of this work.}
         \label{fig:439_mod2}
   \end{figure*}

       \begin{figure*}
   \centerline{\includegraphics[scale=0.39, angle=270]{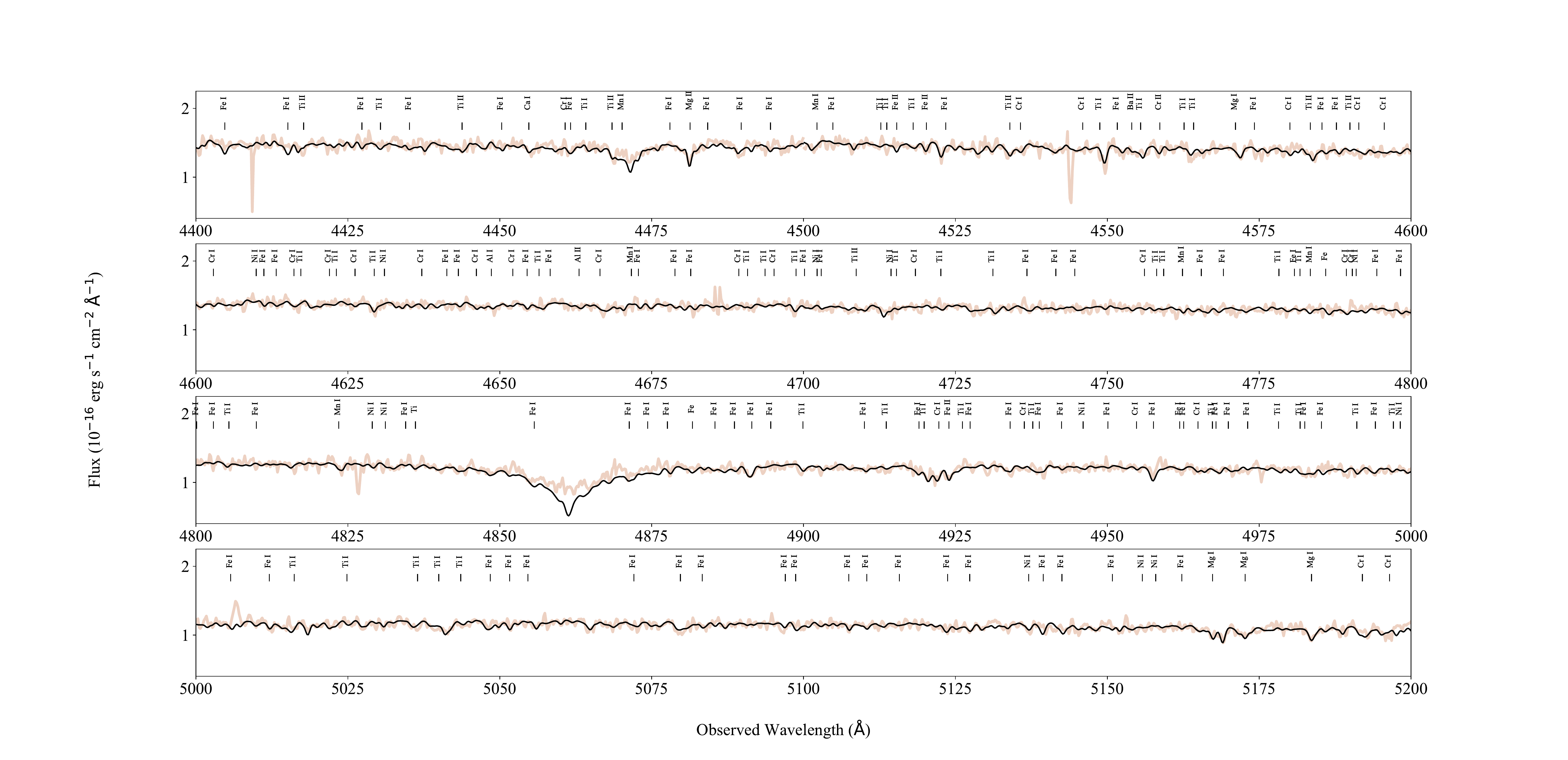}}
      \caption{Example synthesis fits for NGC 1313-463 (UVB X-Shooter coverage). In black we show the best-fitting model spectra. These models use the final abundances obtained as part of this work.}
         \label{fig:463_mod}
   \end{figure*}

       \begin{figure*}
   \centerline{\includegraphics[scale=0.39, angle=270]{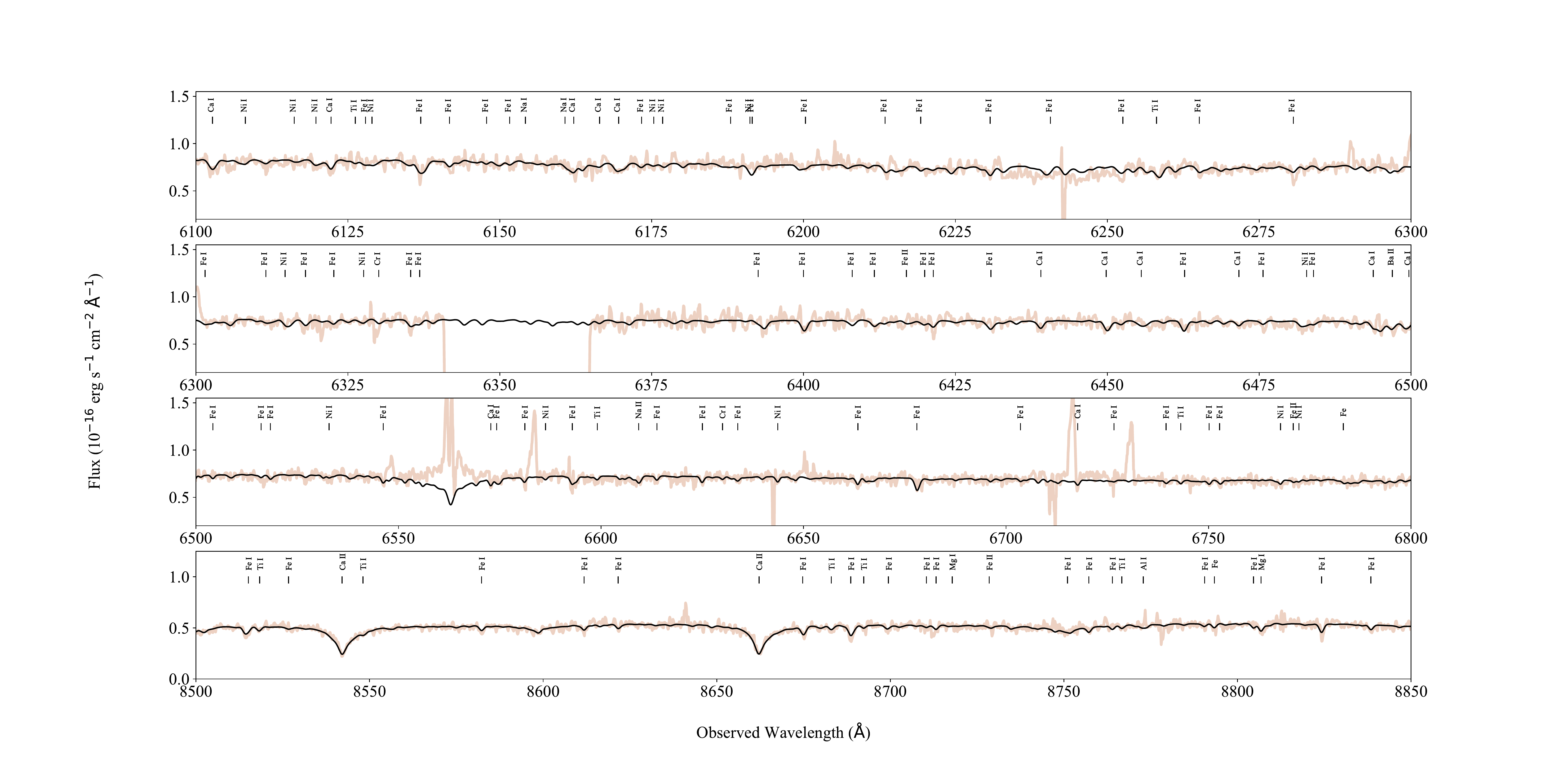}}
      \caption{Example synthesis fits for NGC 1313-463 (VIS X-Shooter coverage). In black we show the best-fitting model spectra. These models use the final abundances obtained as part of this work.}
         \label{fig:463_mod2}
   \end{figure*}

       \begin{figure*}
   \centerline{\includegraphics[scale=0.46, angle=270]{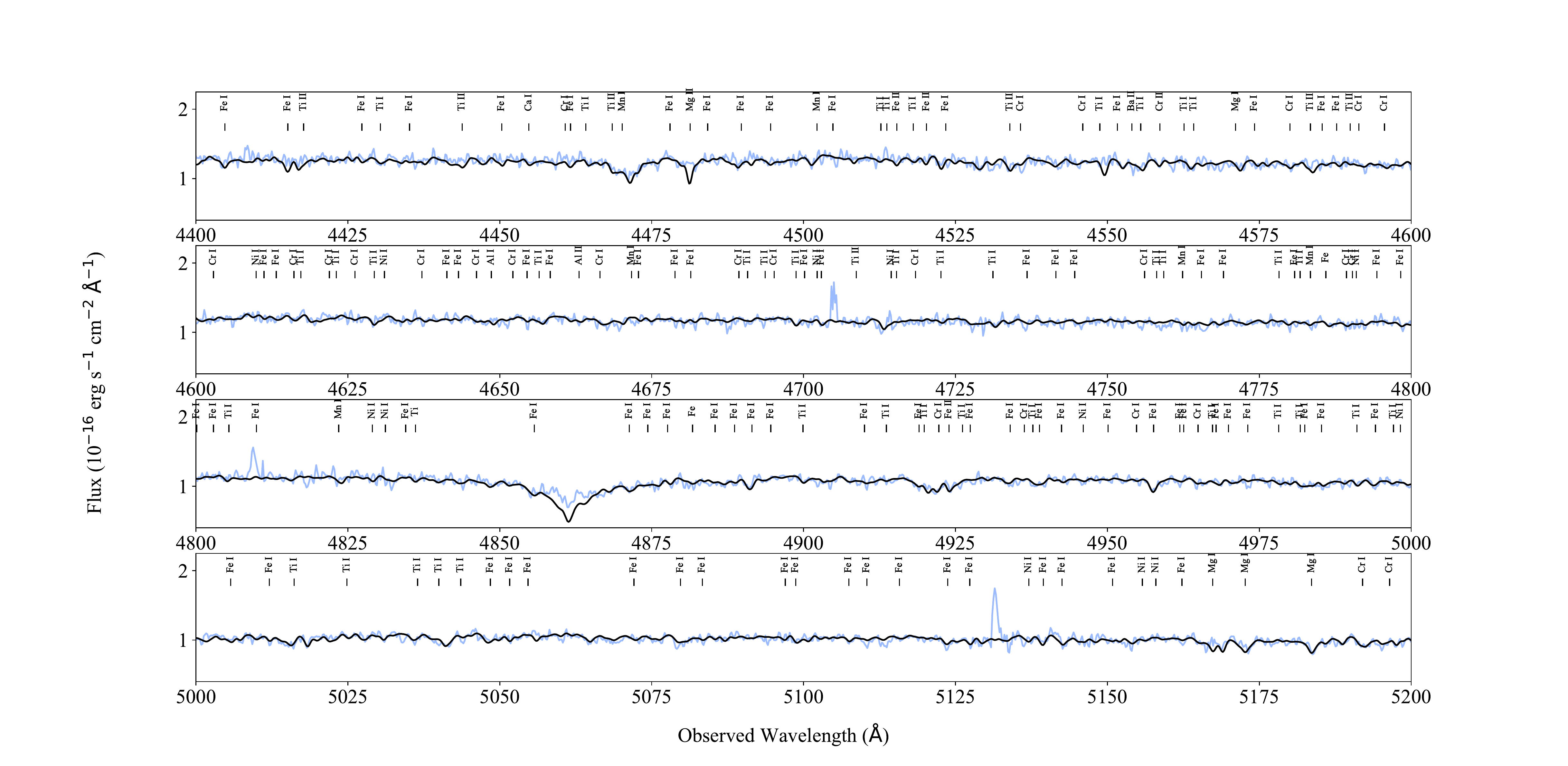}}
      \caption{Example synthesis fits for NGC 1313-503 (UVB X-Shooter coverage). In black we show the best-fitting model spectra. These models use the final abundances obtained as part of this work.}
         \label{fig:503_mod}
   \end{figure*}

       \begin{figure*}
   \centerline{\includegraphics[scale=0.46, angle=270]{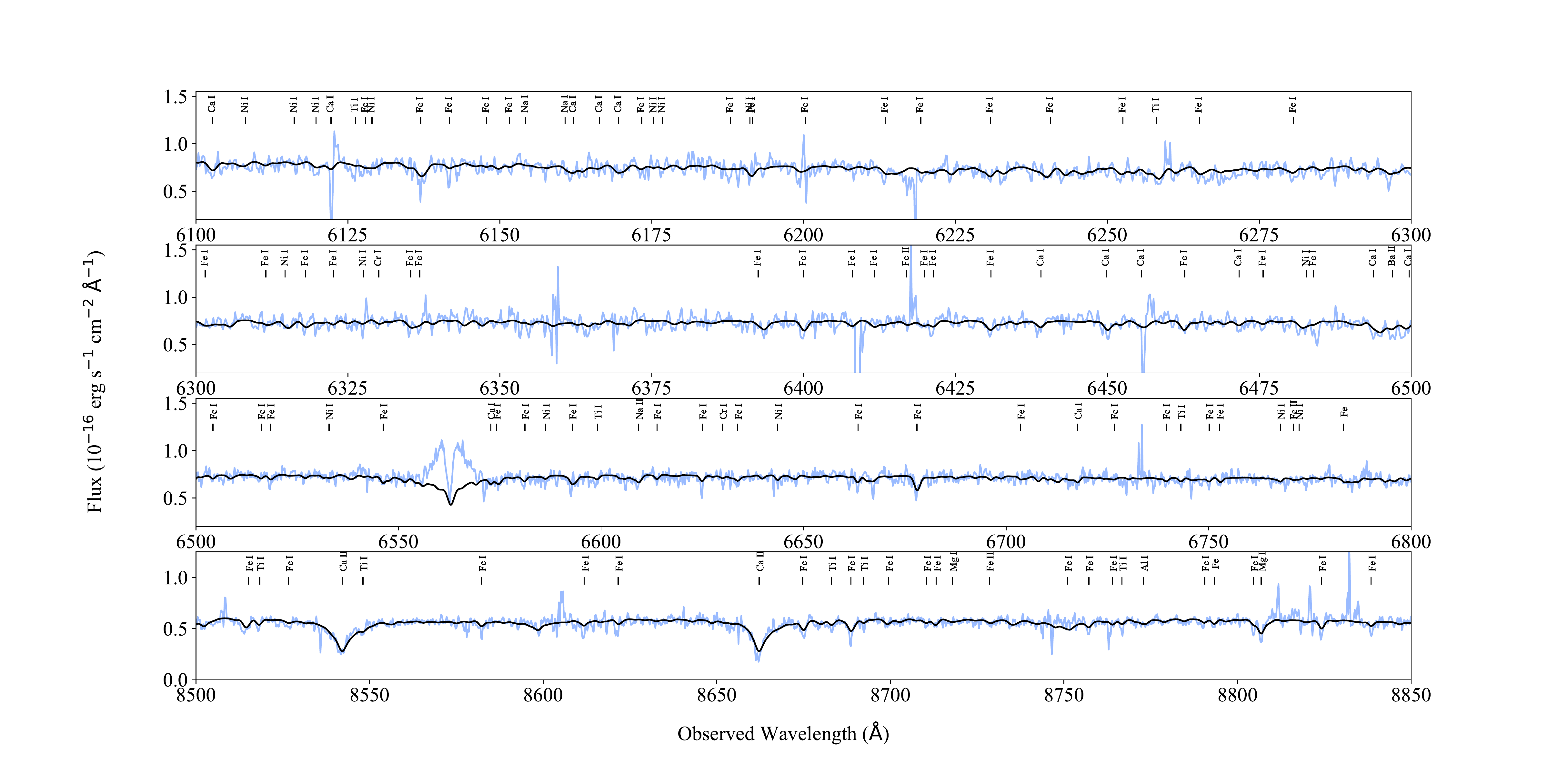}}
      \caption{Example synthesis fits for NGC 1313-503 (VIS X-Shooter coverage). In black we show the best-fitting model spectra. These models use the final abundances obtained as part of this work.}
         \label{fig:503_mod2}
   \end{figure*}

       \begin{figure*}
   \centerline{\includegraphics[scale=0.39, angle=270]{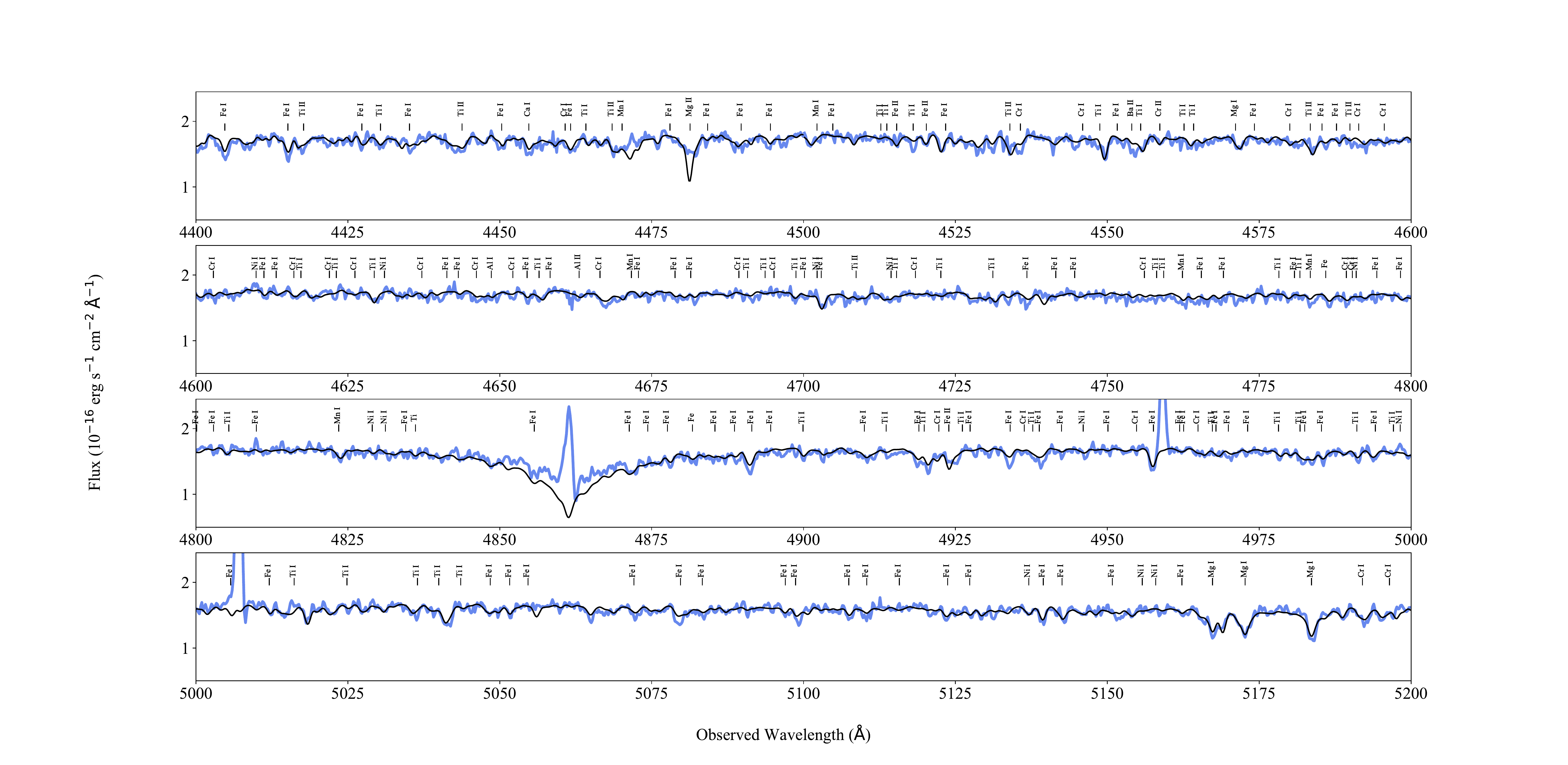}}
      \caption{Example synthesis fits for NGC 1313-505 (UVB X-Shooter coverage). In black we show the best-fitting model spectra. These models use the final abundances obtained as part of this work.}
         \label{fig:505_mod}
   \end{figure*}

       \begin{figure*}
   \centerline{\includegraphics[scale=0.39, angle=270]{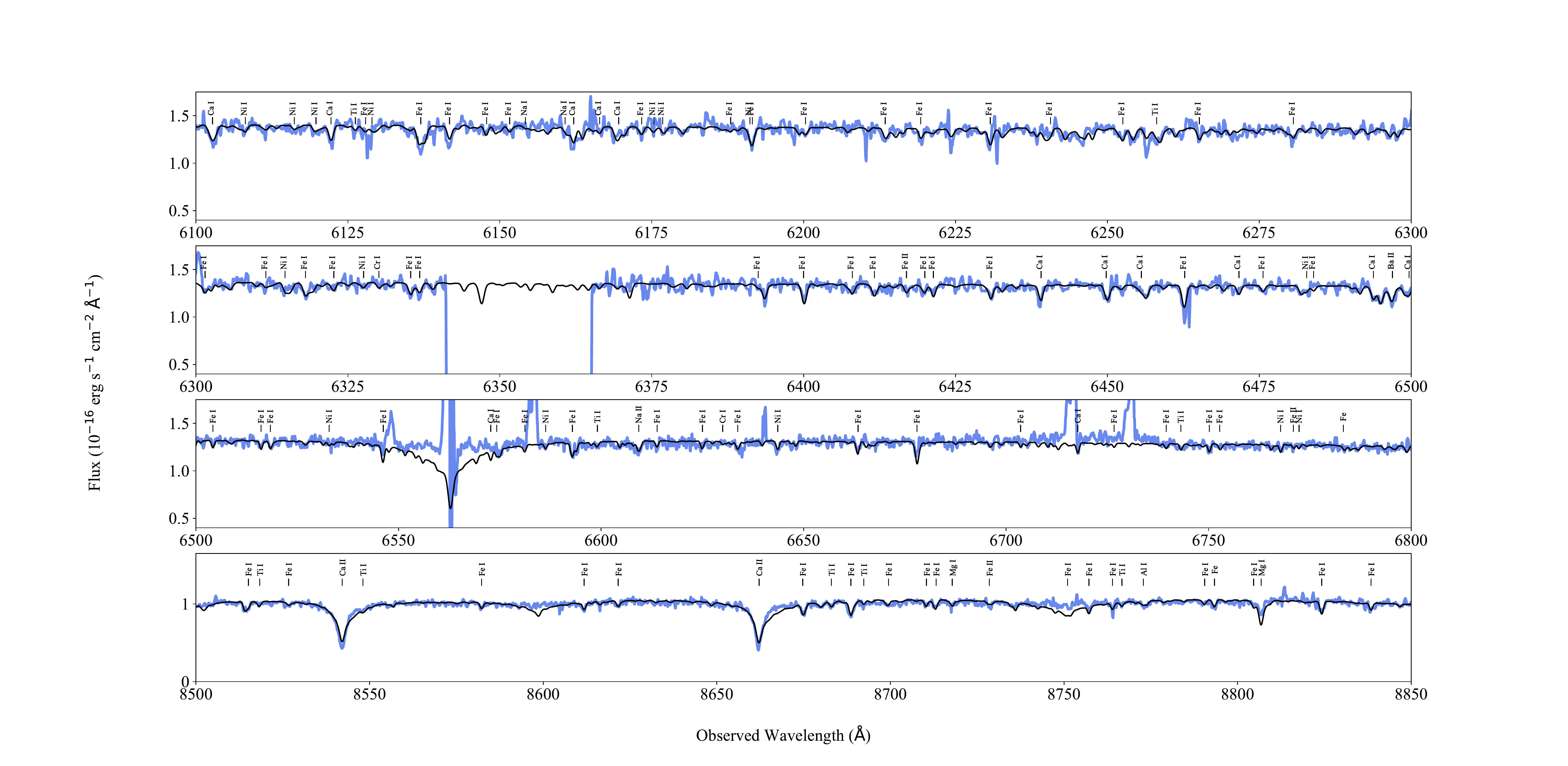}}
      \caption{Example synthesis fits for NGC 1313-505 (VIS X-Shooter coverage). In black we show the best-fitting model spectra. These models use the final abundances obtained as part of this work.}
         \label{fig:505_mod2}
   \end{figure*}

\acknowledgments

%

\vspace{5mm}
\facilities{}

\bibliographystyle{aasjournal}
\bibliography{ngc1313} 


\end{document}